\documentclass[
amsmath,
amssymb,
aps, 
pra,
twocolumn, 
superscriptaddress,
nobalancelastpage,
floatfix,
longbibliography
]{revtex4-2}
\usepackage{mathtools}     
\usepackage{bm}            
\usepackage{graphicx}      
\usepackage{dcolumn}       
\usepackage{xcolor}
\usepackage[colorlinks]{hyperref}
\usepackage{cleveref}

\newcommand{\spaceIndex}[1]{\text{\tiny #1}}

\begin{document}
\author{Mads Brøndum Carlsen}
\author{Lars Bojer Madsen}
\affiliation{Department of Physics and Astronomy, Aarhus University, 8000 Aarhus C, Denmark}

\title{Modulation of electron wave packets by scattering on time-harmonic potentials}


\begin{abstract}
The coherent interaction between free electrons and optical near-fields enables the active modulation of electron wave packets, a mechanism central to photon-induced near-field electron microscopy (PINEM). While existing theories effectively describe these interactions at high kinetic energies, the growing interest in low-energy ultrafast electron microscopy demands frameworks that explicitly account for finite wave packet geometries and recoil effects. In this paper, we develop a rigorous 3D quantum scattering theory for electron wave packets interacting with time-periodic potentials, capturing the case of optical near-field interaction. By mapping the time-dependent dynamics into an extended Floquet space, we formally connect the modulation process to time-independent multi-channel scattering. We evaluate the resulting scattering amplitudes using both an exact R-matrix approach and a multi-channel eikonal approximation. The latter analytical approach recovers PINEM-like probabilities weighted by the wave packet's transverse profile. Application of the theory to an oscillating potential demonstrates the generation of distinct energy sidebands, revealing that the modulation strength is sensitive to the transverse focusing of the incident electron pulse, underlining the importance of a fully 3D treatment. 
\end{abstract}

\maketitle

\section{Introduction}
The ability to control and shape free-electron wave packets has opened new frontiers in ultrafast electron microscopy. Within the last decades, the development of photon-induced near-field electron microscopy (PINEM) \cite{barwick_photon-induced_2009, park_photon-induced_2010,garcia_de_abajo_multiphoton_2010}, has sparked broad interest in coherent interactions between free electrons and photons. In PINEM a fast free electron interacts with the electromagnetic near-field of a nano-structure. The localization of the near-field allows energy exchange between the projectile and field, imprinting a coherent phase modulation on the electronic wave function. Studies have explored the application of such interactions, ranging from the shaping and control over the electronic wave packet \cite{feist_quantum_2015,vanacore_attosecond_2018,giulio_free-electron_2020}, allowing, e.g., formation of attosecond pulse trains \cite{baum_attosecond_2007,kozak_ponderomotive_2018,morimoto_diffraction_2018}, to creation of free-electron qubits \cite{reinhardt_free-electron_2021}, cat-states \cite{dahan_creation_2023} and applications within quantum optics \cite{ruimy_free-electron_2025}.

Most of the studies regarding free-electron interactions assume high kinetic energies, often hundreds of keV. Here, the momentum can be treated almost like a constant parameter, and the non-recoil approximation is applicable. In this approximation, only momentum transfer in the direction of incoming momentum is appreciable, enabling both analytical manipulation and well-defined phase imprints on the longitudinal wave function of the electron. Only recently did studies in the low-energy regime begin \cite{chahshouri_stimulated_2026}. This regime is interesting due to the stronger interaction between target and projectile electrons and the possibility for recoil effects, as has been illustrated numerically using a coupled Maxwell-Schrödinger equation solver \cite{talebi_strong_2020}. Experimental studies on slow electrons have been performed using ultrafast point-projection electron microscopy (UPEM) \cite{vogelsang_observing_2018, hergert_probing_2021}, demonstrating the coupling between optical near-fields and free-electron wave packets of 100 eV and sub $50$ fs duration \cite{woste_ultrafast_2023}. These experimental studies have been accompanied by a theoretical study \cite{hergert_strong_2021}, generalizing the PINEM theory to slower projectiles. 

A few studies have investigated the effect of scattering with modulated electron pulses. For instance, using a semiclassical description of the free electron, it was proposed that a resonance condition could occur when the projectile was modulated at frequencies matching transitions in the target \cite{gover_free-electron--bound-electron_2020}. This caused a debate \cite{de_abajo_comment_2021}, with further studies \cite{zhang_quantum_2021,zhang_quantum_2022, garcia_de_abajo_optical_2021} showing that an effect was indeed present when the target was in a coherent superposition. While these studies rely on effectively 1D models, full 3D scattering theories for wave packets have also been used to investigate scattering with fs Gaussian wave packets \cite{shao_imaging_2013, shao_time-resolved_2013, shao_energy-resolved_2017}, demonstrating the ability to resolve the coherent evolution of an atom in a superposition of states. More recently, such a theory was applied to scattering with modulated wave packets, confirming an effect of the modulation  \cite{morimoto_coherent_2021}; however, the expression for the modulated pulses in that work still relied on the non-recoil approximation. 

Currently, the theoretical work based on 3D wave packet formulations of quantum scattering theory, such as Refs.~\cite{shao_imaging_2013, morimoto_scattering_2024}, has not yet incorporated the modulation of wave packets. The purpose of this paper is to bridge that gap. We consider wave packet scattering on a time-periodic potential, playing the role of the modulating near-field of PINEM, similar to Ref.~\cite{hergert_strong_2021}. The time-periodicity allows treatment in an extended Floquet space, allowing for a time-independent description as a multi-channel scattering problem. The main theoretical work is to connect this description to the actual time-dependent wave packet scattering. The presented theory is beyond the non-recoil approximation, and will be applied to short ($\sim20$ fs), low-energy wave packets of $100$~eV. From the general framework, we further illustrate how a multi-channel eikonal approximation yields equations similar to those of PINEM theory.

The paper is outlined as follows. The scattering theory is derived in Sec.~\ref{sec:theory_1}. Section~\ref{sec:theory_2} presents two approaches for calculating scattering amplitudes, specifically the exact R-matrix method and the eikonal approximation. Section~\ref{sec:results} presents results for wave packet scattering on a radial oscillating potential. Section~\ref{sec:conclusion_outlook} draws the conclusions. For completes a set of appendices is included. Appendix \ref{sec:app_time_dep_scatt}-\ref{sec:asymp_int_trick} contains calculations supporting the scattering theory.  Appendix~\ref{sec:app_r_mat_method} contains details on the R-matrix method. Atomic units (a.u.) are used unless otherwise stated.

\section{Theory - Formulation of scattering theory} \label{sec:theory_1}
In this section, we present a scattering theory for electron wave packets subject to a time-periodic potential. Although primarily motivated by optical near-field interactions, which can be effectively modeled via a scalar potential in the low-energy regime \cite{hergert_strong_2021}, the mathematical framework derived here is general and valid for any time periodic short-range potential. 

The derivations are outlined as follows: In Sec.~\ref{sec:floquet_theory} we briefly cover the essential results from Floquet theory. Section~\ref{sec:time_independent_scatt_theory} then derives results for a time-independent plane wave scattering theory in an extended-Floquet space. These results will then be linked to an explicit time-dependent formulation in Sec.~\ref{sec:Time_dep_scattering}, containing the main results of this section. Finally, Sec.~\ref{sec:psi_k_asymp_diff_prop} considers the asymptotic wave function and various differential probability distributions. 

\subsection{Floquet extended space formalism} \label{sec:floquet_theory}
Throughout the paper, we assume that the Hamiltonian for the scattering system is time periodic $\hat{H}(t+T) = \hat{H}(t)$, where $T=2\pi/\omega$ is the period and $\omega$ the angular frequency of the time-varying potential. Specifically we take 
\begin{equation} \label{time_dep_pot_scatt_hamilton}
	\hat{H}_\spaceIndex{A}(t) = \hat{H}_{\spaceIndex{A}0} + \hat{V}_\spaceIndex{A}(t) = \hat{H}_{\spaceIndex{A} 0}  + \sum_m \hat{V}_{\spaceIndex{A}m} e^{im\omega t},
\end{equation}
where $\hat{H}_{\spaceIndex{A}0}$ is the Hamiltonian of the free projectile electron, and $\hat{V}_{\spaceIndex{A}m}$ are the Fourier components of the scattering potential, $\hat{V}_\spaceIndex{A}$. The use of subscript `A' will be explained shortly. From Floquet theory, we know the eigenstates of a time-periodic Hamiltonian can be written \cite{joachain_atoms_2011}
\begin{equation} \label{Floquet_quasistates}
	|\Psi_E(t) \rangle_\spaceIndex{A} = e^{-iEt} | \psi_E(t)\rangle_\spaceIndex{A} = e^{-iEt} \sum_n e^{in\omega t} |\psi_{E,n}\rangle_\spaceIndex{A},
\end{equation}
where $E$ is the Floquet `quasi' energy and $|\psi_E(t)\rangle\spaceIndex{A}$ is a $T$-periodic state. Furthermore, in the last equality, we have utilized the periodicity to expand the states in a Fourier series. In the Floquet extended space formalism \cite{poertner_validity_2020,holthaus_floquet_2015,sambe_steady_1973}, we extend our original Hilbert space (we call it `atomic' space, `A'), by the infinite dimensional space spanned by the orthonormal time-periodic functions used for Fourier decomposition, $|n\rangle_\spaceIndex{F}$, where $n\in \mathbb{Z}$ (we call it `Fourier' space, `F'). In this formalism, the states in atomic space, Eq.~\eqref{Floquet_quasistates}, can be written in terms of corresponding states in the extended space as
\begin{equation} \label{Floquet_quasi_state}
	|\psi_E(t)\rangle_\spaceIndex{A} = \sum_n e^{in\omega t} {}_\spaceIndex{F}\langle n |\psi_E\rangle_\spaceIndex{FA},
\end{equation}
where the subscripts indicate which space the states belong to, `FA' indicating the product space $\text{F} \otimes \text{A}$. The importance of this equation is to create a well-defined link from the extended space back to atomic space. 

Following the Supplementary Material of Ref.~\cite{poertner_validity_2020}, the Schrödinger equation for the states Eq.~\eqref{Floquet_quasi_state} is written as an eigenvalue problem 
\begin{equation} \label{Floquet_eqn}
	\hat{H}_\spaceIndex{FA} |\psi_E\rangle_\spaceIndex{FA} = E |\psi_E \rangle_\spaceIndex{FA},
\end{equation}
where the extended space Hamiltonian takes the form 
\begin{equation}
	\hat{H}_\spaceIndex{FA} = \sum_n n\omega |n\rangle\langle n|_\spaceIndex{F} + \sum_m \hat{S}_\spaceIndex{F}(m) \tilde{H}_\spaceIndex{A}(m).
\end{equation}
Here $\hat{S}_\spaceIndex{F}$ is a shift operator in Fourier space 
\begin{equation}
	\hat{S}_\spaceIndex{F}(m) = \sum_n |n+m\rangle \langle n|_\spaceIndex{F},
\end{equation}
and $\tilde{H}_\spaceIndex{A}(m)$ is the $m$'th Fourier component of the atomic space Hamiltonian. Last, one can, in the extended space, obtain the following expression for the time-evolution operator
\begin{equation} \label{U_Floquet}
	\hat{U}_\spaceIndex{A}(t,t') = \sum_n e^{in\omega t} {}_\spaceIndex{F}\langle n | e^{-i\hat{H}_\spaceIndex{FA}(t-t')} |0\rangle_\spaceIndex{F},
\end{equation}
similar to that of Shirley \cite{shirley_solution_1965}.

\subsection{Time-independent scattering in extended space} \label{sec:time_independent_scatt_theory}
The main point in working in the extended space is to remove the explicit time-dependence from the Hamiltonian. In this sense, it is a natural starting point for the development of time-independent scattering theory. We follow standard formulations \cite{joachain_collision}, deviating only in the use of an extended space to handle the periodic time-dependence. 

We start by writing the Hamiltonian of Eq.~\eqref{time_dep_pot_scatt_hamilton} in extended space as 
\begin{equation}
	\hat{H}_\spaceIndex{FA} = \underbrace{\sum_n n\omega |n\rangle\langle n|_\spaceIndex{F} + \hat{H}_{\spaceIndex{A}0}}_{\hat{H}_{\spaceIndex{FA}0}} + \underbrace{\sum_{m} \hat{S}_\spaceIndex{F}(m) \hat {V}_{\spaceIndex{A}m}}_{\hat{V}_\spaceIndex{FA}}
\end{equation}
The $\hat{H}_{\spaceIndex{FA}0}$ is the free Hamiltonian for the electron in the extended space. From this, we define the Green's operator 
\begin{equation}
	\hat{G}(E) = \frac{1}{E-\hat{H}_{\spaceIndex{FA}0}}.
\end{equation}
This operator is diagonal in the product basis $|n\rangle_\spaceIndex{F} |\bm k\rangle_\spaceIndex{A}$, where $|\bm k\rangle_\spaceIndex{A}$ is a momentum eigenstate: 
\begin{equation}
	G_{nm}(\bm k, \bm k') = {}_\spaceIndex{F}\langle n| {}_\spaceIndex{A}\langle \bm k| \hat{G}(E) |m\rangle_\spaceIndex{F} |\bm k' \rangle_\spaceIndex{A} = \frac{\delta(\bm k - \bm k')\delta_{n,m}}{E - k^2/2-n\omega}.
\end{equation}
The Green's function corresponding to outgoing spherical waves can be obtained by shifting the pole off the real axis, by introducing a small shift $i\eta$
\begin{equation} \label{Green_func_k_reg}
	G^{(+)}_{nm}(\bm k, \bm k') = \frac{\delta(\bm k - \bm k')\delta_{n,m}}{E - k^2/2-n\omega + i\eta},
\end{equation}
as long as the limiting procedure $\lim\limits_{\eta\rightarrow 0}$ is implied at the end of a calculation. From this, we find the Green's function corresponding to outgoing waves in atomic position space as 
\begin{equation} \label{G_photon_channel}
	G^{(+)}_{nm}(\bm r, \bm r') = - \frac{\delta_{n,m}}{2\pi} \frac{e^{ik_n|\bm r - \bm r'|}}{|\bm r - \bm r'|},
\end{equation}
where $k_n=\sqrt{2(E-n\omega)}$. 

Using the Green's function, we obtain time-independent scattering solutions to the Schrödinger equation for energy $E$ as 
\begin{equation} \label{LS_floquet}
	| \psi^{(+)}\rangle_\spaceIndex{FA} = |\psi_0\rangle_\spaceIndex{FA} + \hat{G}^{(+)}(E) \hat{V}_\spaceIndex{FA} | \psi^{(+)}\rangle_\spaceIndex{FA}, 
\end{equation}
where $|\psi_0\rangle_\spaceIndex{FA}$ is an eigenstate of $\hat{H}_{\spaceIndex{FA}0}$ with energy $E$. In the following, we assume an initial state only occupied in the 0\textsuperscript{th} Fourier channel, $|\psi_0\rangle_\spaceIndex{FA} = |0\rangle_\spaceIndex{F} | \bm k_i\rangle_\spaceIndex{A}$. Expanding the solution in Eq.~\eqref{LS_floquet} in Fourier states as 
\begin{equation} \label{psi_expand_fourier_channels}
	|\psi^{(+)}\rangle_\spaceIndex{FA} = \sum_n |n\rangle_\spaceIndex{F} |\psi^{(+)}_n\rangle_\spaceIndex{A},
\end{equation}
and projecting onto $|n\rangle_\spaceIndex{F}$ we find the asymptotic representation in real space, $\psi_n(\bm r) = \langle \bm r | \psi_n^{(+)}\rangle_\spaceIndex{A}$, as 
\begin{equation} \label{real_space_scatt_wf_asymp}
	\psi^{(+)}_n(\bm r) \xrightarrow[r \to \infty]{} \frac{1}{(2\pi)^{3/2}} \left( e^{i \bm k_i \cdot \bm r}\delta_{n,0} + f_n(\bm k_n, \bm k_i)\frac{e^{i\bm k_n r}}{r} \right),
\end{equation}
where the scattering amplitude for the $n$'th Fourier channel takes the form
\begin{equation} 
	f_n(\bm k_n, \bm k_i) 
	= 
	-(2\pi)^2 \sum_m {}_\spaceIndex{A}\langle \bm k_n | \hat{V}_{nm} |\psi^{(+)}_m\rangle_\spaceIndex{A} \label{scatt_amp_vs_Tmat},
\end{equation} 
with matrix elements $\hat{V}_{nm} = {}_\spaceIndex{F}\langle n| \hat{V}_\spaceIndex{FA} | m \rangle_\spaceIndex{F}$ and final momentum $\bm k_n = k_n \hat{\bm r}$. Overall, we see from Eq.~\eqref{real_space_scatt_wf_asymp} that the periodic time-dependence of the scattering potential leads to a time-independent multi-channel scattering problem in the extended space.

In the following section, it will prove useful to obtain our scattering state, $|\psi^{(+)}\rangle_\spaceIndex{FA}$ in terms of $|\psi_0\rangle_\spaceIndex{FA}$ directly. For this, we separate the final scattering state in terms of the incoming state plus a scattered part 
\begin{equation} \label{psi_0_plus_psi_scatt}
	|\psi^{(+)}\rangle_\spaceIndex{FA} = |\psi_0\rangle_\spaceIndex{FA} + | \psi_\text{scatt}\rangle_\spaceIndex{FA}
\end{equation}
Inserting Eq.~\eqref{psi_0_plus_psi_scatt} in the Floquet equation, Eq.~\eqref{Floquet_eqn}, leads to 
\begin{equation}
 (E-\hat{H}_\spaceIndex{FA}) |\psi_\text{scatt}\rangle_\spaceIndex{FA} = \hat{V}_\spaceIndex{FA}|\psi_0\rangle_\spaceIndex{FA},
\end{equation}
allowing us to the state in Eq.~\eqref{psi_0_plus_psi_scatt} as
\begin{equation} \label{tot_LS_extended_space}
	|\psi^{(+)}\rangle_\spaceIndex{FA} = |\psi_0 \rangle_\spaceIndex{FA} + \hat{\mathcal{G}}^{(+)}(E) \hat{V}_\spaceIndex{FA} | \psi_0\rangle_\spaceIndex{FA}, 
\end{equation}
where $\hat{\mathcal{G}}^{(+)}(E) = (E + i\eta - \hat{H}_\spaceIndex{FA})^{-1}$ is the Green's operator for the total extended space Hamiltonian. Again, the small parameter $\eta$ is introduced to ensure the correct boundary conditions, and the limit $\lim\limits_{\eta \rightarrow 0}$ is implied.

\subsection{Time-dependent scattering} \label{sec:Time_dep_scattering}
In an explicitly time-dependent formulation of scattering theory we consider the scattering state $|\psi(t)\rangle_\spaceIndex{A}$ through time evolution from an asymptotically free `in'-state \cite{goldberger_collision, newton_scattering}
\begin{equation} \label{initial_time_dep_scatt_state}
	|\psi(t)\rangle_\spaceIndex{A} = \lim\limits_{t_0 \to -\infty} \hat{U}_\spaceIndex{A}(t,t_0) \hat{U}_{\spaceIndex{A}0}(t_0,t) |\psi_\text{in}(t)\rangle_\spaceIndex{A}.
\end{equation}
Here $\hat{U}_\spaceIndex{A}(t,t_0)$ is the time-evolution operator from $t_0$ to $t$, corresponding to the full (time-dependent) Hamiltonian $\hat{H}_\spaceIndex{A}(t)$, while $\hat{U}_{\spaceIndex{A}0}$ is the time evolution operator for $\hat{H}_{\spaceIndex{A}0}$ only, corresponding to free motion of the projectile electron. 

In the following, we consider scattering with a wave packet having a confined width in real space, with the in-state generally written through its momentum space wave function, $\psi_0(\bm k)$, as  
\begin{equation} \label{psi_in_time_formalism}
	|\psi_\text{in}\rangle_\spaceIndex{A} = \int d\bm k \, \psi_0(\bm k) |\bm k\rangle_\spaceIndex{A}.
\end{equation}
We note that a momentum state representation is the natural choice, as the momentum states are the eigenstates of the free atomic Hamiltonian, $\hat{H}_{\spaceIndex{A}0}$. Furthermore, we assume a finite extent of the scattering potential. This, together with the finite extent of the wave packet, means that there must exist some time $t_c$ for which
\begin{equation} \label{wp_finite_condition}
	\hat{V}_\spaceIndex{A}(t) |\psi_\text{in}(t)\rangle_\spaceIndex{A} = 0, \quad  t<t_c,
\end{equation}
meaning a finite spatial overlap of potential and wave packet is not present at asymptotic times. Following arguments from Ref.~\cite{goldberger_collision}, Eq.~\eqref{wp_finite_condition} allow us to handle the asymptotic time limit in Eq.~\eqref{initial_time_dep_scatt_state}, see also Appendix~\ref{sec:app_time_dep_scatt}. 

We proceed by expanding the time evolution operator $\hat{U}_\spaceIndex{A}$ in Eq.~\eqref{initial_time_dep_scatt_state} by an integral representation \cite{joachain_atoms_2011} 
\begin{equation} \label{psi_time_evolv_expr}
	|\psi(t)\rangle_\spaceIndex{A} =  |\psi_\text{in}(t)\rangle_\spaceIndex{A} - \lim\limits_{t_0 \to -\infty} i \int_{t_0}^t dt' \hat{U}_\spaceIndex{A}(t,t')\hat{V}_\spaceIndex{A}(t') |\psi_\text{in}(t')\rangle_\spaceIndex{A}.
\end{equation}
In order to carry out the integral, $\hat{U}_\spaceIndex{A}$ must have a suitable simple form. In time-independent scattering, $\hat{U}_\spaceIndex{A}$ is an exponential of time. In the current Floquet setting, we can obtain a similar case by employing Eq.~\eqref{U_Floquet}. Following the steps in Appendix \ref{sec:app_time_dep_scatt} we obtain for the scattering state: 
\begin{multline}
	|\psi(t)\rangle_\spaceIndex{A} = 
	 \int d\bm k \, \psi_0(\bm k) e^{-iE_kt} \Big( |\bm k \rangle_\spaceIndex{A} 
	 \\ +
	 \sum_n e^{in\omega t} \sum_m e^{im\omega t} \hat{\mathcal{G}}^{(+)}_{n0}(E_k - m\omega) \hat{V}_{\spaceIndex{A}m} | \bm k \rangle_\spaceIndex{A} \Big), \label{time_dep_final}
\end{multline}
where $\hat{\mathcal{G}}^{(+)}_{nm} = {}_\spaceIndex{F}\langle n | \hat{\mathcal{G}}^{(+)} |m \rangle_\spaceIndex{F}$ is the  matrix elements of the total Green's operator between Fourier space states. 

The next step is to connect the result of Eq.~\eqref{time_dep_final} to the time-independent theory of Sec.~\ref{sec:time_independent_scatt_theory}. As the time-independent scattering state, $|\psi^{(+)}\rangle_\spaceIndex{FA}$, is a solution to the Floquet equation Eq.~\eqref{Floquet_eqn}, it follows from Eq.~\eqref{Floquet_quasi_state} that its representation in pure atomic space has the form
\begin{equation}
	| \psi^{(+)}(t)\rangle_\spaceIndex{A} = \sum_n e^{in\omega t} {}_\spaceIndex{F}\langle n| \psi^{(+)}\rangle_\spaceIndex{FA}.
\end{equation}
Again considering the initial state $|\psi_0\rangle_\spaceIndex{FA} = |0\rangle_\spaceIndex{F} |\bm k\rangle_\spaceIndex{A}$, we find, using Eq.~\eqref{tot_LS_extended_space},
\begin{equation} \label{tot_LS_atomic_space}
	| \psi^{(+)}(t)\rangle_\spaceIndex{A} = |\bm k\rangle_\spaceIndex{A} + \sum_n e^{in\omega t} \sum_m \hat{\mathcal{G}}^{(+)}_{nm}(E) \hat{V}_{\spaceIndex{A}m} | \bm k\rangle_\spaceIndex{A}.
\end{equation}
This is almost the $\bm k$ components of Eq.~\eqref{time_dep_final}. To enable direct comparison, we clearly need to shift $m \to 0$ in the Green's function. To this end, note 
\begin{equation}
 \hat{S}^\dagger_\spaceIndex{F}(p)\hat{H}_\spaceIndex{FA} \hat{S}_\spaceIndex{F}(p) = \hat{H}_\spaceIndex{FA} + p\omega
\end{equation}
such that 
\begin{equation}
	\hat{S}^\dagger_\spaceIndex{F}(p) \hat{\mathcal{G}}^{(+)}(E) \hat{S}_\spaceIndex{F}(p) =  \hat{\mathcal{G}}^{(+)}(E-p\omega).
\end{equation}
For the matrix element in Eq.~\eqref{tot_LS_atomic_space}, we can therefore write 
\begin{equation}
	\hat{\mathcal{G}}^{(+)}_{nm}(E)  = \hat{\mathcal{G}}^{(+)}_{n-m,0}(E-m\omega), 
\end{equation}
and Eq.~\eqref{tot_LS_atomic_space} becomes, under a relabeling of the $n$ summation index, 
\begin{equation}
	| \psi^{(+)}(t)\rangle_\spaceIndex{A} = |\bm k\rangle_\spaceIndex{A} + \sum_{n,m} e^{i(n+m)\omega t} \hat{\mathcal{G}}^{(+)}_{n0}(E-m\omega) \hat{V}_{\spaceIndex{A}m} | \bm k\rangle_\spaceIndex{A}.
\end{equation}
Now, comparing with Eq.~\eqref{time_dep_final}, we see that the total time-dependent wave packet can be written as 
\begin{align} \label{floquet_scatt_wavefunc}
	|\psi(t)\rangle_\spaceIndex{A} 
	&=
	\int d \bm k \, \psi_0(\bm k) e^{-iE_k t} | \psi^{(+)}_{\bm k}(t)\rangle_\spaceIndex{A} \nonumber 
	\\ &= 
	\int d \bm k \, \psi_0(\bm k) e^{-iE_k t} \sum_n e^{in\omega t} |\psi_{\bm k, n}^{(+)}\rangle_\spaceIndex{FA},
\end{align}
where we have added the subscript $\bm k$ to explicitly point out that it is the scattering state corresponding to incoming momentum $\bm k$, and expanded the scattering state in Fourier components, similar to Eq.~\eqref{psi_expand_fourier_channels}.  

We have hereby established the connection between the full time-dependent scattering state and the time-independent scattering states in the extended space. Perhaps unsurprisingly, we see that the effect is to map the initial momentum states in the incoming wave packet to their different Floquet state components, $|\bm k \rangle_\spaceIndex{A} \to | \psi^{(+)}_{\bm k,n}\rangle_\spaceIndex{FA}$.

\subsection{Differential probabilities} \label{sec:psi_k_asymp_diff_prop}
When calculating experimental observables, we are mostly interested in the asymptotic wave function in momentum space. In Appendix \ref{sec:app_asymp_state_momentum} we show that in the $t \rightarrow \infty$ limit, the momentum representation of Eq.~\eqref{floquet_scatt_wavefunc}, evaluated at a final momentum $\bm k_f$ of interest, takes the form 
\begin{multline} \label{scatt_state_momentum}
	\psi(\bm k_f) \xrightarrow[t \to \infty]{} e^{-iE_{k_f}t} \left[ \psi_0(\bm k_f) \vphantom{\sum_n}\right. 
	\\
	+ \left. \frac{i}{2\pi} \sum_n \int d \Omega_k k_{in} \psi_0(k_{in}\hat{\bm k}) f_n(\bm k_f, k_{in}\hat{\bm k}) \right],
\end{multline}
where $k_{in}=\sqrt{2E_{k_{in}}}$ is the momentum corresponding to initial energy $E_{k_{in}} = E_{k_f} + n\omega$, such that the final energy is $E_{k_f}$ under exchange of $n$ Fourier energy quanta, $n\omega$. 

With the above asymptotic wave function, we can compute various values of interest. Here, we consider the differential probability in momentum (energy) 
\begin{equation} \label{tot_differential_scatt_prop_k}
	\frac{dP}{k_f^2 dk_f} = \int d\Omega_{k_f} |\psi(\bm k_f)|^2 = P_D(k_f) + P_S(k_f) + P_I(k_f).
\end{equation}
In the last part, we have separated the differential probability into its three components: The differential probability corresponding directly to the incoming wave packet, without scattering taking place,
\begin{equation} \label{diff_prop_k_direct}
	P_D(k_f) = \int d\Omega_{k_f}|\psi_0(\bm k_f)|^2,
\end{equation}
the differential probability originating from scattering only 
\begin{multline} \label{diff_prop_k_scatt}
	P_S(k_f) = \frac{1}{4\pi^2} \int d\Omega_{k_f} 
	\\
	\times \left| \sum_n k_{in} \int d\Omega_k \psi_0(k_{in}\hat{\bm k}) f_n(\bm k_f, k_{in}\hat{\bm k}) \right|^2,
\end{multline} 
and the differential probability corresponding to interference between the incoming and scattered parts 
\begin{multline} \label{diff_prop_k_inter}
	P_I(k_f) = \frac{1}{\pi} \int d\Omega_{k_f} \operatorname{Re}\left\{ i \psi_0^*(\bm k_f) \sum_n  k_{in} \vphantom{\sum_n \int} \right. 
	\\
	\times \left. \int d\Omega_k \psi_0(k_{in}\hat{\bm k}) f_n(\bm k_f, k_{in}\hat{\bm k}) \right\}.
\end{multline}
In Sec.~\eqref{sec:R_mat_results}, these expressions are used to calculate the differential probability Eq.~\eqref{tot_differential_scatt_prop_k} for scattering on an oscillating potential. The results will be presented as a function of energy, i.e.,
\begin{equation} \label{dP_dE}
	\frac{dP}{dE_f} = k_f \frac{dP}{k_f^2dk_f}.
\end{equation}
First, however, we need to determine the scattering amplitudes, entering the above expressions, and this determination is the subject of the following section.


\section{Theory - Calculation of scattering amplitudes} \label{sec:theory_2}
In this section, we consider two methods for practical calculation of the scattering amplitudes $f_n$, Eq.~\eqref{scatt_amp_vs_Tmat}. As an exact numerical approach, we first consider the R-matrix approach. Then we consider an approximate method, namely the semi-classical eikonal approximation, which leads to equations for the scattering wave functions similar to those of PINEM theory. The often used Born approximation \cite{joachain_collision} is not considered, due to the fact that the scattering potential in Eq.~\eqref{Floquet_eqn} or Eq.~\eqref{LS_floquet} is also what allows exchange of Fourier energy quanta. In order to allow exchange of multiple Fourier energy quanta we would thus have to go to increasingly higher Born orders, which is impractical for numerical evaluation. We note, however, that such an approach has been used within the non-recoil approximation in early theoretical work on PINEM \cite{garcia_de_abajo_multiphoton_2010}.  

\subsection{R-matrix method}
Our aim is to solve the Floquet equation, Eq.~\eqref{Floquet_eqn}, for a scattering state 
\begin{equation} \label{rmat_floquet_eqn}
	\hat{H}_\spaceIndex{FA} |\psi^{(+)}\rangle_\spaceIndex{FA} = E|\psi^{(+)}\rangle_\spaceIndex{FA}.
\end{equation}
If we expand our states in Fourier channels like in Eq.~\eqref{psi_expand_fourier_channels}, and project down on Fourier state $|n\rangle_F$, we obtain
\begin{equation} \label{coupled_floquet_eqns} 
	(\hat{H}_{\spaceIndex{A}0} + n\omega)|\psi_n^{(+)}\rangle_\spaceIndex{A} + \sum_{m} \hat{V}_{\spaceIndex{A}m} |\psi_{n-m}^{(+)}\rangle_\spaceIndex{A}
	=
	E|\psi_n^{(+)}\rangle_\spaceIndex{A}.
\end{equation}
Clearly, this is a set of coupled equations for the different Fourier channel functions, completely analogous to the closed-coupling equations found in time-independent multi-channel scattering theory \cite{burke_rmat_2011, friedrich_theoretical_2017}. 

A widely used approach for solving these equations in scattering theory is the R-matrix method \cite{burke_rmat_2011}. The method separates real space into an inner and outer part. In the inner part, the radial part of the coupled equations is solved by diagonalization, where after linear combinations of eigenstates are matched to the outer, known asymptotic solutions. The interested reader can find the detailed steps in Appendix~\ref{sec:app_r_mat_method}. In brief, the resulting scattering amplitudes are
\begin{multline} \label{scatt_amp_S_mat_relation}
	f_{n}(k_n\hat{\bm k}_f, \bm k_i) = \frac{2\pi i}{\sqrt{k_i k_n}} \sum_{\ell m} Y_\ell^m(\hat{\bm k_f}) \sum_{\ell_i m_i} i^{\ell_i - \ell} Y_{\ell_i}^{m_i}(\hat{\bm k_i})^* 
	\\
	\times \left(\delta_{n,0}\delta_{\ell, \ell_i}\delta_{m,m_i} - S_{n\ell m, 0 \ell_i m_i}\right),
\end{multline}
where $\bm k_i$ is the initial momentum, $\hat{\bm k}_f$ the direction of final momentum with $k_n=\sqrt{k_i^2 -2n\omega}$ its length, and $S_{n\ell m, n_i\ell_im_i}$ the S-matrix element describing scattering from Fourier channel $n_i$, angular quantum number $\ell_i$ and magnetic quantum number $m_i$ to the corresponding final set $(n,\ell,m)$. Since we always assume an initial Fourier channel of 0, we have taken $n_i=0$. 

Let us consider a few interesting special cases of Eq.~\eqref{scatt_amp_S_mat_relation}. Take, for example, the situation where the scattering potential is spherically symmetric, i.e., the $\ell$ and $m$ quantum numbers are conserved. This means $S_{n\ell m, 0 \ell_i m_i} = S_{n,0}^{(\ell m)}\delta_{\ell,\ell_i}\delta_{m,m_i}$. From Eq.~\eqref{scatt_amp_S_mat_relation} we, in this case, obtain 
\begin{equation} \label{scatt_amp_spherical}
	f_{n}(k_n\hat{\bm k}_f, \bm k_i) = \frac{2\pi i}{\sqrt{k_i k_n}} \sum_{\ell m} Y_\ell^m(\hat{\bm k}_f) Y_{\ell}^{m}(\hat{\bm k_i})^* \left(\delta_{n,0} - S^{(\ell m)}_{n,0} \right).
\end{equation}
Further specializing to the case $\hat{\bm k_i} = \hat{ \bm z}$ (standard setting in plane wave scattering), 
\begin{equation}
	f_{n}(k_n\hat{\bm k}_f, \bm k_i) = \frac{i}{\sqrt{k_i k_n}} \sum_\ell Y_\ell^0(\hat{\bm k}_f) \sqrt{\pi (2\ell+1)} \left(\delta_{n,0} - S^{(\ell 0)}_{n,0} \right),
\end{equation}
which agrees with Eq.~(4.313) in Ref.~\cite{friedrich_theoretical_2017}. Eq.~\eqref{scatt_amp_spherical} will be applied in calculations in Sec.~\ref{sec:R_mat_results}.

\subsection{Eikonal approximation}
In this section, we consider an approximate way of obtaining the scattering amplitudes, using the well-known eikonal approximation \cite{joachain_collision}. This approximation is valid for fast electrons (small wave length compared to potential range), where the non-recoil approximation can be applied, and thus unsurprisingly it yields equations similar to those of PINEM theory. We further note that some of the early work on PINEM theory \cite{garcia_de_abajo_multiphoton_2010} actually applied a Green's function linearization, similar to that used in the eikonal approximation, discussed below. We follow a formulation from scattering theory as outlined in Ref.~\cite{joachain_collision} and generalize to our multi-channel case. 

Our starting point is the Lippmann-Schwinger equation Eq.~\eqref{LS_floquet}, which, for $|\psi_0\rangle_\spaceIndex{FA} = |0\rangle_\spaceIndex{F} | \bm k_0\rangle_\spaceIndex{A}$, leads to the following equations for the channel functions 
\begin{multline} \label{Eikonal_start_LS_eqn}
	\psi_n^{(+)}(\bm r) = \frac{1}{(2\pi)^{2/3}}e^{i \bm k_0 \cdot \bm r}\delta_{n,0} + 
	\\
	\sum_\ell \int d \bm r' G_{n\ell}^{(+)}(\bm r, \bm r') \sum_m \hat{V}_{\spaceIndex{A}m}(\bm r') \psi_{\ell - m}^{(+)}(\bm r').
\end{multline}
The eikonal approximation is based on a linearization of the Green's function. To perform this, we assume that $\bm k_0 = k_0 \hat{\bm z}$, and that the incident momentum is large. Consider the Green's function in the following representation, directly obtainable from Eq.~\eqref{Green_func_k_reg}, 
\begin{equation} \label{greens_func_space_k_int}
	G_{n\ell}^{(+)}(\bm r, \bm r') = \frac{-1}{(2\pi)^3} \int d \bm k \frac{e^{i\bm k \cdot (\bm r-\bm r')}\delta_{n\ell}}{k^2/2 - k_0^2/2 + n\omega -i\eta}.
\end{equation}
Shifting variables to the momentum change $\bm p = \bm k - \bm k_0$, we can write the following terms of the denominator as $k^2-k_0^2 = p^2 + k_0^2 + 2\bm p\cdot \bm k_0 \sim k_0^2 + 2\bm p\cdot \bm k_0$. In the last equality, we neglected terms of order $p^2$, under the assumption that the momentum exchange is small for large $k_0$. Back in Eq.~\eqref{greens_func_space_k_int}, the integrals along $p_x$ and $p_y$ trivially yield $\delta$-functions, while the integral along $p_z$ is handled with contour integration 
\begin{multline}
	G_{n\ell}^{(+)}(\bm r, \bm r') \sim  \frac{-1}{(2\pi)^3} e^{i\bm k_0 \cdot \bm R} \int d\bm p \frac{e^{i\bm p \cdot \bm R}}{\bm p \cdot \bm k_0 + n\omega -i\eta} \delta_{n\ell} 
	\\ = -i\delta(R_x) \delta(R_y) \theta(R_z) \frac{1}{k_0}e^{i(k_0 - n\omega/k_0)R_z - \eta R_z} \delta_{n\ell}, \label{eikonal_approx_greens_func}
\end{multline}
where $\bm R = \bm r - \bm r'$. As seen in the phase of the exponential, the approximate Green's function only allows exchange of momentum along the incoming ($z$) direction of the projectile, in units of 
\begin{equation}
	\delta_k =  \frac{\omega}{k_0}.
\end{equation}
Inspired by this fact, and common ansätze in PINEM theory \cite{garcia_de_abajo_multiphoton_2010, giulio_free-electron_2020}, we separate out the dominant plane wave part of our channel functions by writing 
\begin{equation} \label{eikonal_ansatz}
	\psi_n^{(+)}(\bm r) = \frac{1}{(2\pi)^{3/2}} e^{i k_0 z - in\delta_k z} \phi_n(\bm r),
\end{equation}
where $\phi_n$ is now to be determined. Inserting Eq.~\eqref{eikonal_ansatz} in Eq.~\eqref{Eikonal_start_LS_eqn}, together with the Green's function Eq.~\eqref{eikonal_approx_greens_func}, we obtain the following integral equations 
\begin{multline} \label{eikonal_int}
	\phi_n(\bm r) \sim \delta_{n,0} - \frac{i}{k_0} \sum_m \int_{-\infty}^z dz' \, e^{imz'\delta_k} 
	\\
	\times V_{\spaceIndex{A}m}(z', \bm r_\perp) \phi_{n-m}(z', \bm r_\perp), 
\end{multline}
where we write the spatial coordinates as $\bm r = (z,\bm r_\perp)$, with $\bm r_ \perp$ being a vector in the $xy$-plane. By taking the derivative wrt. $z$, Eq.~\eqref{eikonal_int} can be turned into the following differential equations 
\begin{equation} \label{eikonal_diff_eqn}
	\frac{\partial \phi_n (\bm r)}{\partial z} = -\frac{i}{k_0} \sum_{m} e^{imz\delta_k} V_{\spaceIndex{A}m}(\bm r) \phi_{n-m}(\bm r).
\end{equation}
In the case of $m = \pm1$ only, these differential equations match those of PINEM-theory, see, e.g., Eq.~(3) in \cite{garcia_de_abajo_electron_2016}. 

\subsubsection{Scattering amplitude from eikonal wave function}
The eikonal wave function Eq.~\eqref{eikonal_ansatz} is not a valid scattering wave function in the asymptotic region. This is clear from Eq.~\eqref{eikonal_diff_eqn} that shows that $\phi_n(\bm r)$ is constant along $z$ for values of $r$ outside the range of the potential $V_{\spaceIndex{A}m}(\bm r)$, meaning Eq.~\eqref{eikonal_ansatz} does not fulfill the required spherical wave condition of Eq.~\eqref{real_space_scatt_wf_asymp}. Instead, we use the eikonal wavefunction in the region where it is approximately valid, to determine the scattering amplitude through Eq.~\eqref{scatt_amp_vs_Tmat}. This definition only requires the scattering wave function in the finite region of support of the scattering potential. 
\begin{multline} \label{scatt_amp_eikonal_full}
	f^{E}_n(\bm k_n, \bm k_0) = - \sqrt{2\pi} \int d\bm r' \, e^{-i\bm k_n \cdot \bm r'} \sum_m V_{nm}(\bm r') \psi^{(+)}_m(\bm r')
	\\ = 
	-\frac{1}{2\pi} \sum_m \int d \bm r' \, e^{i(\bm k_0 - \bm k_n)\cdot \bm r' - iz(n-m)\delta_k}V_{\spaceIndex{A}m}(\bm r') \phi_{n-m}(\bm r'),
\end{multline}
using $\hat{V}_{nm} = \hat{V}_{\spaceIndex{A},n-m}$. The integral in Eq.~\eqref{scatt_amp_eikonal_full} is very similar to the one in Eq.~\eqref{eikonal_int}, suggesting simplifications might be possible. In the spirit of the eikonal approximation, we take our integration of the $z'$ variable to be along the line of $\bm k_0=k_0\hat{\bm z}$, i.e., we write $\bm r' = \bm r_\perp' + z' \hat{\bm k}_0$. Using this separation 
\begin{equation} \label{r_dot_k_expr}
	\bm r' \cdot (\bm k_0 - \bm k_n) = \bm{\Delta k}_{n\perp} \cdot \bm r_\perp' + z' (k_0 - k_n \hat{\bm k}_n \cdot \hat{\bm k}_ 0).
\end{equation}
For the high energies where the eikonal approximation is valid, it should be enough to consider only small scattering angles $\theta$, $\hat{\bm k}_n \cdot \hat{\bm k} = \cos(\theta) \sim 1$, and it is also assumed that $\omega$ is small relative to $k_0$, such that $k_n = \sqrt{k_0^2 - 2n\omega} \sim k_0 - n\delta_k$. Therefore Eq.~\eqref{r_dot_k_expr} simplifies as 
\begin{equation}
	\bm r' \cdot (\bm k_0 - \bm k_n) = \bm{\Delta k}_{n\perp} \cdot \bm r_\perp' + z' n\delta_k,
\end{equation}
the clear benefit being that the $z'$ term now cancels a phase in Eq.~\eqref{scatt_amp_eikonal_full}, allowing us to further use Eq.~\eqref{eikonal_diff_eqn} to write
\begin{multline} \label{eikonal_approx_scatt_amp}
	f^{E}_n(\bm k_n, k_0\hat{\bm z}) \sim 
	-\frac{ik_0}{2\pi} \int d\bm r_\perp' \, e^{i \bm r_\perp' \cdot \bm{\Delta k}_{n\perp}} \int dz' \frac{\partial \phi_n(\bm r')}{\partial z'}
	\\= 
	-\frac{ik_0}{2\pi} \int d\bm r_\perp' \, e^{i \bm r_\perp' \cdot \bm{\Delta k}_{n\perp}} \left[\phi_n(\bm r_\perp', z=\infty) - \delta_{n,0} \right].
\end{multline}
This is our final expression for the eikonal scattering amplitude. To assess the angular range of validity, we consider the lowest order of neglected angular terms  
\begin{equation} \label{eikonal_angular_cond}
	\frac{1}{2}z'k_n \theta^2 \sim \frac{1}{2} a_{\text{eff}} k_n \theta^2 \quad \Rightarrow \quad \theta^2 \ll \frac{2}{k_n a_\text{eff}},
\end{equation}
where $a_\text{eff}$ denotes an effective range of the scattering potential. 

\subsubsection{Scattering probability} \label{sec:eikonal_scatt_prop}
With the approximate scattering amplitude obtained as in Eq.~\eqref{eikonal_approx_scatt_amp}, we are now in a position to obtain scattering probabilities. Our goal in this section is to show that under a series of approximations on $f_n^E$, we can obtain scattering probabilities similar to those found in other works on PINEM. The total scattering probability is obtained from Eq.~\eqref{diff_prop_k_scatt} 
\begin{equation} \label{tot_scatt_prop_eikonal_sec}
	P_S = \frac{1}{4\pi^2} \int d \bm k_f \left| \sum_n k_{in} \int d\Omega_k \psi_0(k_{in}\hat{\bm k}) f_n(\bm k_f, k_{in}\hat{\bm k}) \right|^2.
\end{equation}
Since we, in this section, are using an approximate expression for the scattering amplitude, it is clear that the initial wave packet $\psi_0(\bm k)$ must restrict the integral to regions where Eq.~\eqref{eikonal_approx_scatt_amp} remains valid. We thus assume that $\psi_0$ is tightly peaked around an initial energy, $\bm k_0 = k_0 \hat{\bm z}$, such that the factor $\psi(k_{in}\hat{\bm k})$ effectively forces $k_{in} \sim k_0$ under the integral. Furthermore, this energy should be high enough that the first-order expansion 
\begin{equation} \label{linear_k_diff_eikonal}
	k_f = \sqrt{k_{0}^2 - 2n\omega} \sim k_{0} - n\delta_k,
\end{equation}
is satisfied, and we can consider small-angle scattering only, in order for Eq.~\eqref{eikonal_angular_cond} to hold. In this $\delta_k = \omega/k_0 \ll 1$ limit, the different terms in the $n$ sum in Eq.~\eqref{tot_scatt_prop_eikonal_sec} do not overlap and we write 
\begin{multline}
	P_S \sim \sum_n P_S(n) = \sum_n \frac{1}{4\pi^2} k_0^2 
	\\
	\times  \int d \bm k_f \left| \int d\Omega_k \psi_0(k_{in}\hat{\bm k}) f_n(\bm k_f, k_{0}\hat{\bm k}) \right|^2 .
\end{multline}
In order to apply Eq.~\eqref{eikonal_approx_scatt_amp}, we have to consider the trajectories used to evaluate the eikonal wave functions $\phi_n$. In principle, we should perform evaluations for all possible $\hat{\bm k}$. In line with the above approximations, we instead evaluate all $\phi_n$ along the central momentum $k_0\hat{\bm z}$, thus also fixing the $\bm r_\perp$ integration to be in the $xy$-plane. We do, however, keep the $\hat{\bm k}$ dependence in the phase, for reasons that will be clear later. Applying these approximations, we obtain
\begin{multline} \label{temp_ps_eikonal}
	P_S^E(n) \sim \frac{1}{16\pi^4} \int d\bm k_f k_0^4 
	\\
	\times \left| \int d\Omega_k d\bm r_\perp \psi_0(k_{in} \hat{\bm k}) e^{i\bm r_\perp \cdot (k_0 \hat{\bm k} - \bm k_f)} \left[ \phi_n^{k_0\hat{\bm z}}(\bm r_\perp) - \delta_{n,0}\right]\right|^2,
\end{multline} 
where we have added the superscript $E$ to indicate that the eikonal scattering amplitude has now been used. For small angle scattering $k_0\hat{\bm k} \sim k_0 \theta \hat{\bm r}_\perp + k_0 \hat{\bm z} = \bm k_\perp + k_0 \hat{\bm z}$, and we can approximate the angular integral as $d\Omega_k \sim d\bm k_\perp / k_0^2$. Further assuming that the wave packet is separable in $z$ and $xy$ components, we write 
\begin{multline} \label{wavefunc_eikonal_expansion}
	\psi_0(k_{in}\hat{\bm k}) \sim \psi_0(k_{in}\theta \hat{\bm r}_\perp + k_{in}\hat{\bm z}) = \psi_{0\perp}(k_{in}\theta \hat{\bm r}_\perp)\psi_{0\parallel}(k_{in}) 
	\\
	\sim \psi_{0\perp}(k_{0}\theta\hat{\bm r}_\perp)\psi_{0\parallel}(k_{fz} + n\delta_k).
\end{multline}
Using Eq.~\eqref{wavefunc_eikonal_expansion} in Eq.~\eqref{temp_ps_eikonal} we can expand the absolute value and perform the $\bm k_{f\perp}$ integration, yielding a $\delta$-function in $\bm r_\perp$, leaving us with 
\begin{multline}
	P_S^E(n) \sim \frac{1}{4\pi^2} \int_{-\infty}^\infty d k_{fz} \left|\psi_{0\parallel}(k_{fz} + n\delta_k) \right|^2
	\\
	\times \int d\bm r_\perp \left| \int d \bm k_\perp  \psi_{0\perp}(\bm k_{\perp}) e^{i\bm r_\perp \cdot k_0 \hat{\bm k}_\perp } \left[ \phi_n^{k_0\hat{\bm z}}(\bm r_\perp) - \delta_{n,0}\right]\right|^2.
\end{multline}
Assuming each part of the wave packet is normalized, the integral over $k_{fz}$ gives unity, and we are left with 
\begin{equation} \label{final_eikonal_scatt_prop}
	P_S^E(n) \sim \frac{1}{4\pi^2} \int d\bm r_\perp \left| \phi_n^{k_0\hat{\bm z}}(\bm r_\perp) - \delta_{n,0} \right|^2 \rho_\perp(\bm r_\perp),
\end{equation}
where
\begin{equation} \label{eikonal_perp_density}
	\rho_\perp(\bm r_\perp) = \left| \int d \bm k_\perp  \psi_{0\perp}(\bm k_{\perp}) e^{i\bm r_\perp \cdot \bm k_\perp} \right|^2
\end{equation}
is the probability density of the perpendicular part of the wave packet in real space. A similar analysis can be performed for the interference term, yielding 
\begin{equation} \label{final_eikonal_interference}
	P_I^E \sim \frac{1}{2\pi^2} \int d \bm r_\perp \operatorname{Re}\left[\phi_0^{k_0 \hat{\bm z}}(\bm r_\perp) - 1\right] \rho_\perp(\bm r_\perp) .
\end{equation}
The total probability for the electron to be in Fourier channel $n$ post-scattering is then found to be 
\begin{equation} \label{tot_eikonal_channel_prop}
	P^E(n) = \delta_{n,0}  \left( 1 + P_I^E \right) + P_S^E(n),
\end{equation}
where the 1 in the initial channel is the direct scattering probability. This relation will be applied in Sec.~\ref{sec:eikonal_res} in a comparison against the R-matrix results. 

In both Eq.~\eqref{final_eikonal_scatt_prop} and Eq.~\eqref{final_eikonal_interference}, the real space perpendicular probability density, $\rho_\perp(\bm r_\perp)$, acts as a weight function over `impact parameters', $\bm r_\perp$, determining which eikonal trajectories contribute to the total scattering probability. Much of the work on PINEM theory is one dimensional, or effectively so, and thus scattering amplitudes in the PINEM framework are directly related to the eikonal wave functions, $P_S^E \sim \sum_n |\phi_n-\delta_{n,0}|^2$, see, e.g., Refs.~\cite{garcia_de_abajo_multiphoton_2010, feist_quantum_2015}  In our case this would correspond to a wave packet with large transversely width in momentum space (very transverse focused in real space), such that only a single impact parameter $\bm r_\perp$ would contribute in the integrals, Eqs.~\eqref{final_eikonal_scatt_prop}, \eqref{final_eikonal_interference}, above. 

\section{Scattering on a spherical oscillating potential} \label{sec:results}
In this section, we consider a simple case for the application of the theory outlined above, presenting results obtained with both the R-matrix method and the eikonal approximation. We take the oscillating part of the Hamiltonian to be a simple radial potential with finite extent
\begin{equation} \label{explicit_hamiltonian}
	H_\spaceIndex{A}(t) = \frac{\hat{\bm p}^2_\spaceIndex{A}}{2} + V_{\spaceIndex{A}0}(r)\cos(\omega t + \phi).
\end{equation} 
For the calculations presented, the potential takes the form of a Gaussian 
\begin{equation} \label{explicit_scatt_pot}
	V_{\spaceIndex{A}0}(r) = \frac{V_0}{\sqrt{2\pi \sigma_V}} \exp\left( -\frac{r^2}{2\sigma_V^2}\right),
\end{equation}
where $V_0$ is used to control the strength of the potential and $\sigma_V$ its extent in real space. We generally pick $\sigma_V=10$ a.u. (FWHM of 23.5 a.u. $\sim$ 1.2 nm), which we note is small compared to the usual extend of near-fields surrounding nanostructures, see, e.g., Refs.~\cite{park_photon-induced_2010, garcia_de_abajo_multiphoton_2010, hergert_strong_2021}. It should also be pointed out that the above potential cannot actually represent the interaction between an electron and an electromagnetic near-field, since the spherical nature of the potential prohibits exchanges in angular momenta components. For this reason, we refrain from using the word 'photons' when discussing the Fourier energy quanta. The potential is chosen for ease of calculation, while still allowing an illustration of the properties of the theory and an assessment of the accuracy of the eikonal approximation. 

\subsection{Results with R-matrix method} \label{sec:R_mat_results}
For the Hamiltonian in Eq.~\eqref{explicit_hamiltonian}, the scattering amplitude takes the form of Eq.~\eqref{scatt_amp_spherical}. Our goal is to calculate the differential probability Eq.~\eqref{tot_differential_scatt_prop_k}. To simplify the integrals, we expand our initial $\bm k$-space wave function in spherical components 
\begin{equation} \label{init_wp_spherical_seperated}
	\psi_0(\bm k) = g(k) \sum_{\ell, m} c_{\ell m} Y_\ell^m(\hat{\bm k}),
\end{equation}
i.e., we assume a separation of angular and $k$ components. This form covers, e.g., the `$|k|$-Gaussian' wave packet from Ref.~\cite{morimoto_scattering-asymmetry_2025}
\begin{equation} \label{k-gauss}
	\psi_0(\bm k) = N \exp \left(-\frac{(k-k_0)^2}{4\sigma_k^2} \right) \exp\left( -\frac{\sin^2 \theta_k}{4\sigma_\theta^2} \right),
\end{equation}
where $\sigma_k$ controls the width in momentum length (i.e., energy) and $\sigma_\theta$ the angular width of the wave packet. In practical calculations, we determine the $c_{\ell m}$ coefficients of Eq.~\eqref{init_wp_spherical_seperated} by projecting the angular part of Eq.~\eqref{k-gauss} onto spherical harmonics, while the $g(k)$ part is matched directly. Afterwards, we numerically normalize the $k$ and angular components individually.

Using Eq.~\eqref{init_wp_spherical_seperated} and Eq.~\eqref{scatt_amp_spherical} we obtain the following differential scattering probabilities from Eq.~\eqref{diff_prop_k_direct}-\eqref{diff_prop_k_inter}: 
\begin{align}
	P_D(k_f) &= |g(k_f)|^2 \sum_{\ell} |c_{\ell 0}|^2 \label{diff_direct_spherical}
	\\
	P_S(k_f) &= \frac{1}{k_f} \sum_{\ell} \left| c_{\ell 0}  \Gamma_\ell(k_f) \right|^2 \label{diff_scatt_spherical}
	\\ 
	P_I(k_f) &= -2\frac{g(k_f)}{\sqrt{k_f}} \sum_{\ell}|c_{\ell 0}|^2 \operatorname{Re}\left[ \Gamma_\ell(k_f) \right] \label{diff_inter_spherical},
\end{align} 
where we have defined the `weighted' S-matrix elements as 
\begin{equation} \label{gamma_factor}
	\Gamma_\ell(k_f) = \sum_n \sqrt{k_{in}} \left[ \delta_{n,0} - S^{(\ell 0)}_{n0}(k_{in}) \right] g(k_{in}).
\end{equation}

While there, unsurprisingly, is no phase interference of the angular components in these expressions, the S-matrix elements for different Fourier channels can interfere. Whether or not this happens in practice is determined by the $g(k)$ functions, i.e., the width of the wave packet in $k$ (energy). This point will be investigated further in Sec.~\ref{sec:phase_dep}. 

\subsubsection{System parameters} \label{sec:sys_params}
We consider a wave packet scattering on a potential [Eqs.~\eqref{explicit_hamiltonian}, \eqref{explicit_scatt_pot}] with $\sigma_V=10$ a.u., $V_0=4.1$ a.u., $\omega = 0.057$ a.u., and a phase $\phi=0$, unless otherwise stated. For convergence, a total of $10$ Fourier channels was included to either side of the $n=0$ channel (21 total). The R-matrix boundary was placed at $a_0=50$ a.u., and a B-spline basis \cite{bachau_applications_2001, hart_Bspline_1997} of order 6 with a total of 150 splines based on a linear knot sequence was employed for the diagonalization of the inner region, see also Appendix~\ref{sec:app_r_mat_method}. The central energy of the wave packet was chosen to be $E_0 =k_0^2/2 = 3.67$ a.u. ($\sim$ 100 eV). For the wave packet, we will consider momentum widths of $\sigma_k=0.0005$ a.u. and $\sigma_k=0.007$ a.u., and angular widths of $\sigma_\theta=0.05$ a.u. and $\sigma_\theta=0.02$ a.u. 

Let us briefly comment on the specific choices of simulation parameters. The initial energy, $E_0$, is chosen similarly to that used in UPEM \cite{woste_ultrafast_2023}. The widths of the wave packet in momentum, $\sigma_k$, will, in the limit of small $\sigma_\theta$, correspond to the approximate temporal duration of the wave packets as $\tau \sim \sqrt{2\ln(2)}/\sigma_k k_0$. We consider two cases: A `slim' wave packet with $\sigma_k=0.0005$ a.u. ($\tau \sim 21$ fs) which temporally is in the $<50$ fs range of those observed in the UPEM experiment of Ref.~\cite{woste_ultrafast_2023}, and a `wide' wave packet with $\sigma_k=0.007$ a.u. ($\tau\sim 1.5$ fs) with a temporal width less than a period of the oscillating potential ($T=2.7$ fs). The angular width, $\sigma_\theta$, is for small values related to the transversal width of the wave packet, $\sigma_{k_\perp} \sim k_0\sigma_\theta$, and the real space transversal width $\sigma_{r_\perp} \sim 1/2k_0\sigma_\theta$. We again consider two cases: An angular focused wave packet of $\sigma_\theta=0.02$ a.u. ($\sigma_{r_\perp} \sim 9.3$ a.u.), and a less focused of $\sigma_\theta=0.05$ a.u. ($\sigma_{r_\perp} \sim 3.7$ a.u.), such that one wave packet is approximately equal to the width of the potential ($\sigma_V$) at focus, while the other is more narrow.

\subsubsection{Modulation of wave packets} \label{sec:R_mat_res_mod_of_wp}

\begin{figure}
	\includegraphics[width=1\linewidth]{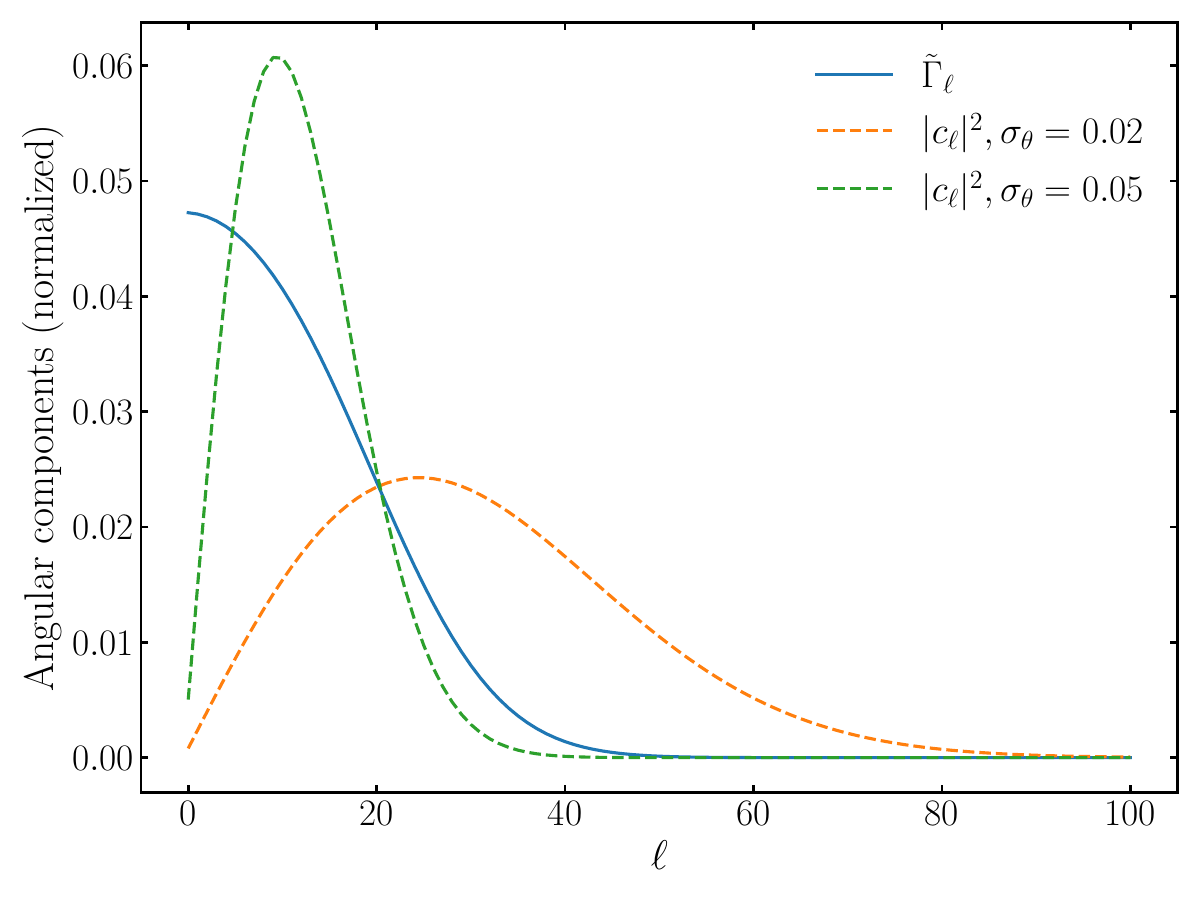}
	\caption{Investigation of the angular components composing the total scattering probability Eq.~\eqref{angular_components}, for system parameters mentioned in the text. The full line shows the momentum integrated potential interaction $\tilde\Gamma_\ell$ [Eq.~\eqref{angular_components}], while the dashed lines show the angular components of wave packets with two different values of $\sigma_\theta$ as indicated by the legend. The results are normalized to enable comparison.}\label{fig:l_components}
\end{figure}

In this section we investigate  the differential scattering probability, Eq.~\eqref{tot_differential_scatt_prop_k} and Eq.~\eqref{dP_dE}, with the different components calculated using Eqs.~\eqref{diff_direct_spherical}--\eqref{diff_inter_spherical}. Before considering probability distributions, we investigate the number of angular components needed in the calculations. As seen in Eq.~\eqref{diff_scatt_spherical}, the maximum required angular component is determined by the angular cutoff of either the wave packet or the potential interaction (through the S-matrix elements). This fact is investigated in Fig.~\ref{fig:l_components}, where the two factors composing the total scattering probability (integral of Eq.~\eqref{diff_scatt_spherical} over final momentum)
\begin{equation} \label{angular_components}
	P_S = \sum_\ell |c_{\ell0}|^2 \int dk_f k_f |\Gamma_\ell(k_f)|^2 = \sum_\ell |c_{\ell0}|^2 \tilde\Gamma_\ell,
\end{equation}
are shown for wave packets with two different angular widths, and $\sigma_k=0.0005$ a.u. As seen, the potential interaction leads to $\ell$ components with significant value up to $\ell \sim 50$. The angular narrow wave packet with $\sigma_\theta=0.02$ a.u. has angular components up to $\ell \sim 80$, and thus in this case the potential interaction ($\tilde\Gamma_\ell$) will be determining the angular cutoff. Opposite the wider wave packet with $\sigma_\theta=0.05$ a.u. has only angular components up to $\ell\sim 40$, meaning it will cut off the angular sum slightly before the potential interaction. 

\begin{figure}
	\includegraphics[width=1\linewidth]{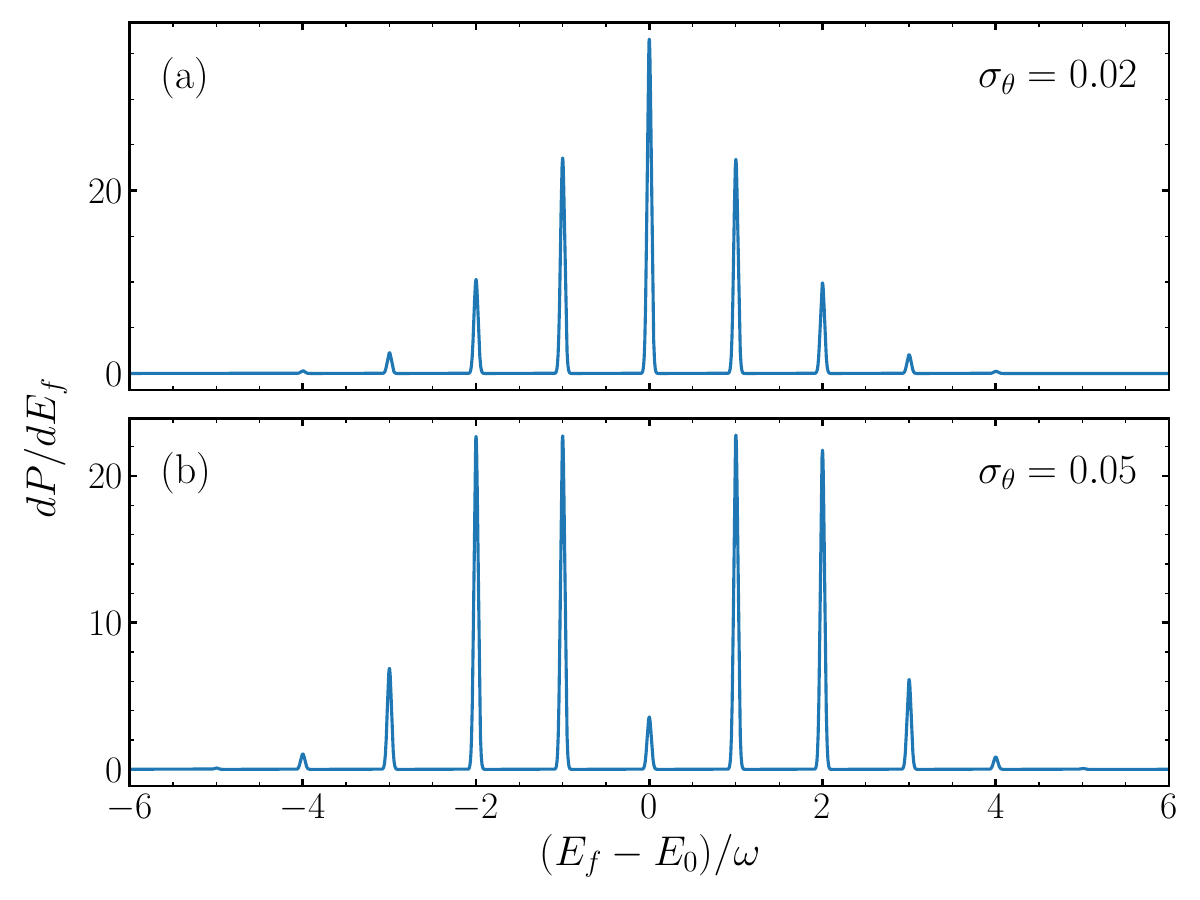}
	\caption{Differential scattering probabilities, Eq.~\eqref{tot_differential_scatt_prop_k} and Eq.~\eqref{dP_dE}, for two wave packets with $\sigma_k=0.0005$ a.u., but different angular widths. The corresponding width in energy is less than $\omega$, thus the Fourier channels are clearly resolved with non-overlapping peaks. (a) Results for $\sigma_\theta=0.02$ a.u. with peaks corresponding to a weak coupling. (b) Results for $\sigma_\theta=0.05$ a.u., with peaks corresponding to strong coupling, with probability largely transferred away from the incoming channel.} 
	\label{fig:k_dist_narrow_wp}
\end{figure}

We now turn to considering the actual differential scattering probability. Fig.~\ref{fig:k_dist_narrow_wp} shows the results for wave packets with the two different angular widths considered above, and a $\sigma_k=0.0005$ a.u. For this small value, the width in energy is less than $\omega$, meaning we see clearly separated peaks for the individual Fourier channels, located at the expected energies of $E_0 + n\omega$. In both panels (a) and (b), a maximum number of 4 energy quanta has been exchanged, and while it is hard to see at the scale of the figure, the peaks located at lower energy are generally a little higher than the corresponding ones at higher energies. Such an asymmetry is expected to be more pronounced at even lower energies of the incoming electron wave packet, where the S-matrix elements have more significant variation over the energy scales considered. 

Fig.~\ref{fig:k_dist_narrow_wp} illustrates the impact of the angular width of the wave packet on the probability distribution. In Fig.~\ref{fig:k_dist_narrow_wp}~(a), where $\sigma_\theta=0.02$ a.u., and the angular cutoff is primarily determined by the potential, the peaks decrease linearly away from the incoming channel. Furthermore, the most prominent peak remains the incoming one. A rather different case is seen in Fig.~\ref{fig:k_dist_narrow_wp}~(b), for $\sigma_\theta=0.05$ a.u. Here, the central peak is found to be almost absent, with the $n=\pm 1$ and the $n=\pm 2$ Fourier channels now being dominant. We thus find that in the case where the angular cutoff is determined by the wave packet, rather than the potential, we are able to transfer more probability to the different Fourier channels. These cases are similar to the weak and strong coupling cases discussed in Ref.~\cite{talebi_strong_2020}, although there the distinction was based on a resonance condition between potential size and oscillation period. An alternative view is provided by considering the approximate transversal widths of the wave packets at focus, as discussed in Sec.~\ref{sec:sys_params}. From there, we see that the `weak' modulation case corresponds to a wave packet spread out over the entire range of the potential, while the `strong' modulation is achieved with a more spatially focused wave packet. The eikonal results presented in Sec.~\ref{sec:eikonal_res} below provide a supplement to this view.

\begin{figure}
	\includegraphics[width=1\linewidth]{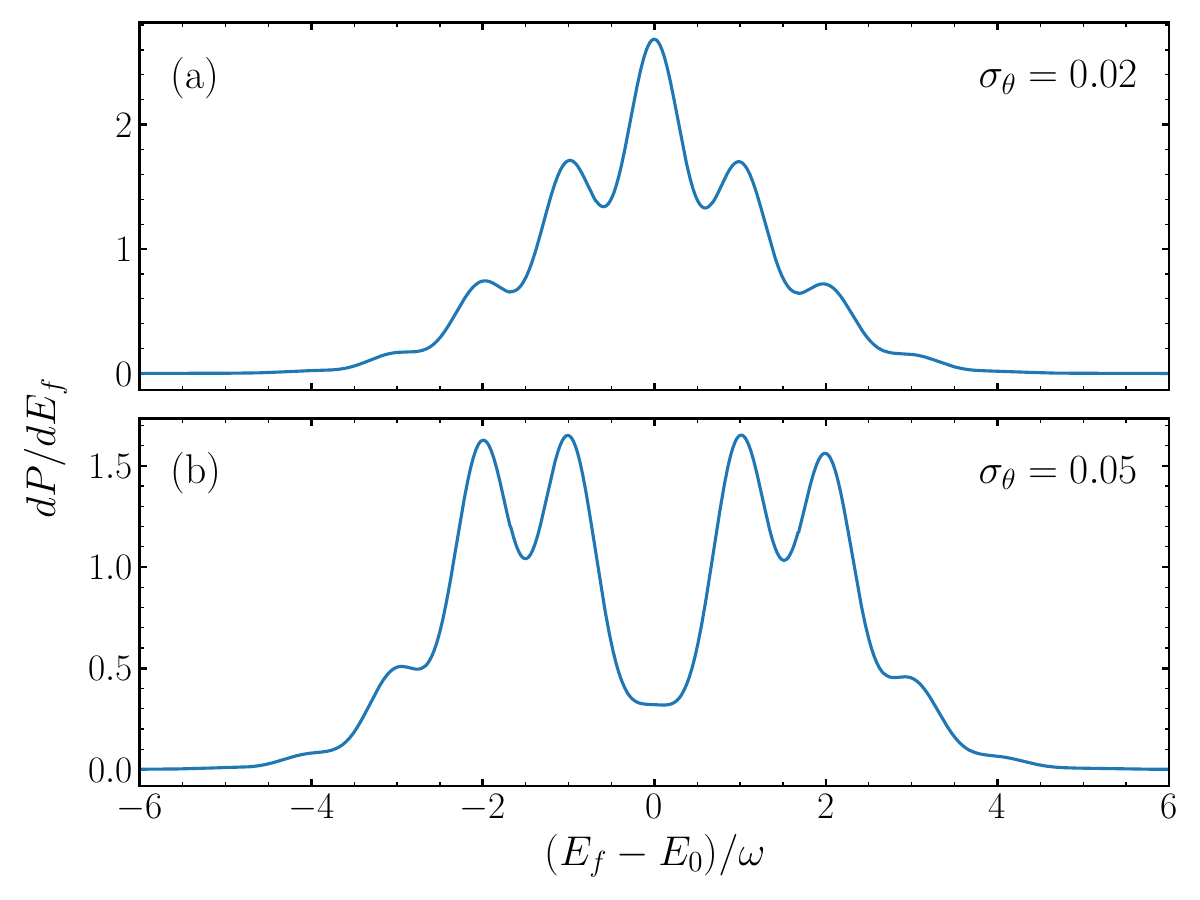}
	\caption{Differential scattering probabilities, Eq.~\eqref{tot_differential_scatt_prop_k} and Eq.~\eqref{dP_dE}, for two wave packets with $\sigma_k=0.007$ a.u., but different angular widths, (a) $\sigma_\theta=0.02$ a.u. and (b) $\sigma_\theta=0.05$ a.u. The corresponding width in energy is such that the Fourier channels overlap, but expect from a energy broadening, the results closely reflect Fig.~\ref{fig:k_dist_narrow_wp}.}
	\label{fig:k_dist_wide_wp}
\end{figure}

Results for a similar set of simulations, but now with a wider wave packet in $k$, $\sigma_k=0.007$ a.u., are shown in Fig.~\ref{fig:k_dist_wide_wp}. In this case, the corresponding width in energy is such that the peaks for the different Fourier channels overlap, and interference is possible. Besides a widening of the peaks, the overall structure of the data is identical to that of Fig.~\ref{fig:k_dist_narrow_wp}. In Sec.~\ref{sec:phase_dep}, we investigate in more detail how the overlap in energy reflects a dependence on the potential phase $\phi$.   

\subsubsection{Scan of potential strength}
\begin{figure*}
	\centering
	\includegraphics[width=1\linewidth]{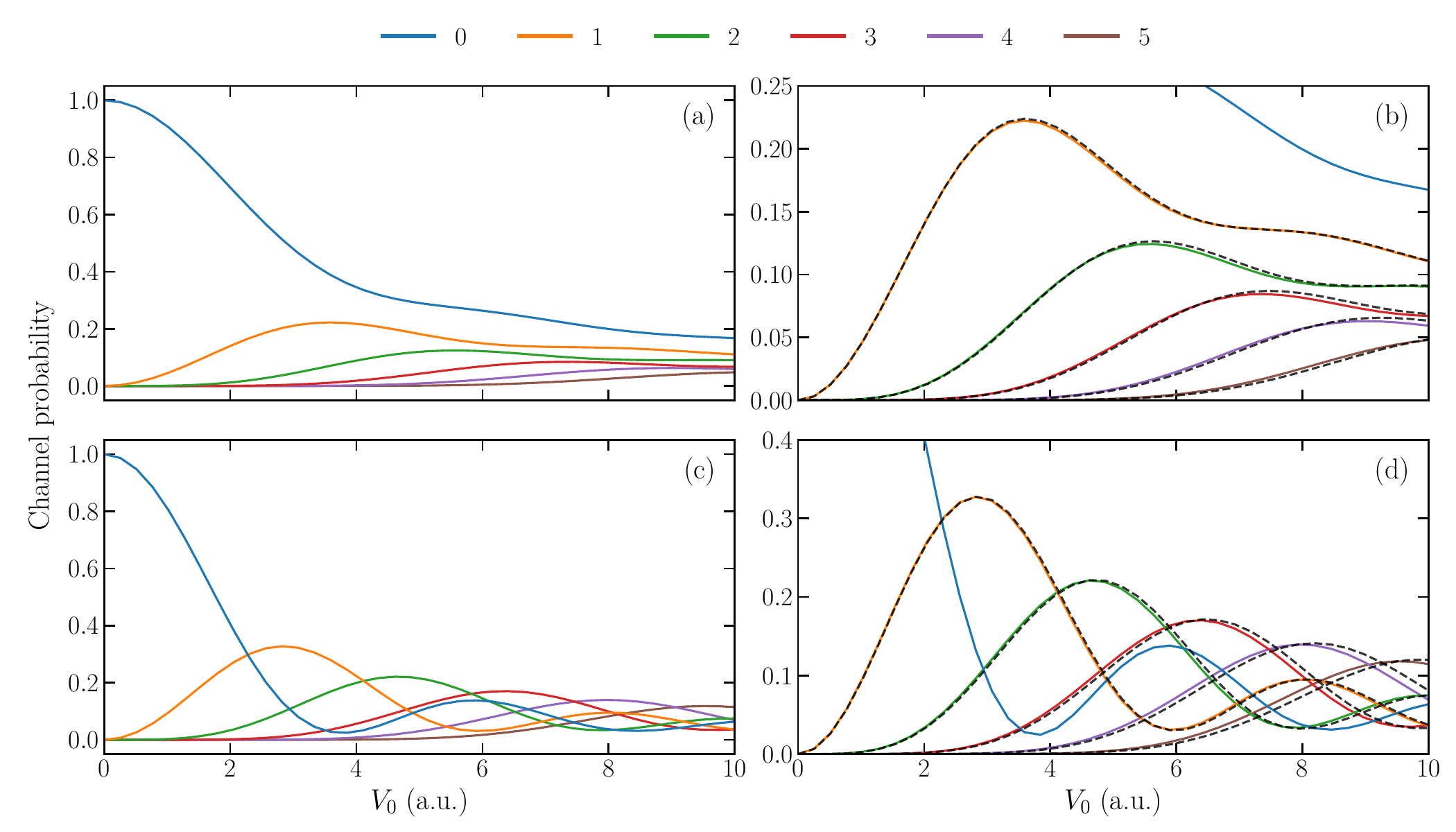}
	\caption{Total channel probabilities for the six first Fourier channels, indicated in the legend, as a function of potential strength, $V_0$. The channel probabilities are found by integrating over the individual peaks in Fig.~\ref{fig:k_dist_narrow_wp}. (a) Channel probabilities for a wave packet with angular width $\sigma_\theta=0.02$ a.u., Fig.~\ref{fig:k_dist_narrow_wp}~(a). (b) Zoom-in of panel (a), with black dashed lines showing data for the negative Fourier channels. (c) Channel probabilities for a wave packet with angular width $\sigma_\theta=0.05$ a.u., Fig.~\ref{fig:k_dist_narrow_wp}~(b). (d) Zoom-in of panel (c), with black dashed lines showing data for the negative Fourier channels.}
	\label{fig:channel_prop_V_scan}
\end{figure*}
In Fig.~\ref{fig:k_dist_narrow_wp}, we saw that a narrow wave packet in momentum space (small value of $\sigma_k$) allowed resolution of the different Fourier channels in the differential scattering probability. By integrating over each peak, we can determine the total probability for the electron to scatter into the different Fourier channels. Such channel probabilities are shown as a function of the potential strength, $V_0$, in Fig.~\ref{fig:channel_prop_V_scan} for the two wave packet configurations of Fig.~\ref{fig:k_dist_narrow_wp}. Panels (a) and (b) show the data for the angular focused wave packet with $\sigma_\theta=0.02$ a.u.. As discussed, the modulation is `weak' in the sense that the incoming channel remains dominant, even when the potential strength is increased significantly. As the potential strength increases, more channels are gradually excited, and as a result, the already open channels gradually decrease in probability.

In Figs.~\ref{fig:channel_prop_V_scan}~(c) and (d), the case for the angular wider wave packet with $\sigma_\theta=0.05$ a.u. is shown. It is seen that at certain potential strengths, the probability for the final state to be in the initial channel is almost zero, corresponding to the strong modulation case seen above. Overall, the channel probabilities here evolve in the Bessel-like fashion that is typical for PINEM modulation \cite{park_photon-induced_2010,feist_quantum_2015}.

In Figs.~\ref{fig:channel_prop_V_scan} (b) and (d), the channel probabilities are plotted for both the positive and corresponding negative Fourier channels. Here it is clear that the channels do not have equal probability, unlike what is found in the eikonal approximation (illustrated below in Fig.~\ref{fig:Eikonal_V0_test}) and PINEM theory. From the figure, it seems that the deviation between the $\pm n$ Fourier channels is more pronounced for higher channel numbers, $n$. This is expected, as the channel pairs are further separated in energy the higher the channel number, meaning the S-matrix elements in Eq.~\eqref{diff_scatt_spherical}, will differ more for the two $\pm n$ channels.

Finally, we note that simulations have been performed for additional values of $\omega$, specifically $\omega=0.1$ a.u. and $\omega=0.024$ a.u. The overall behavior of the channel probabilities reported in Fig.~\ref{fig:channel_prop_V_scan} remains the same, with slight changes in overall values. One tendency that can be observed is that the R-matrix results for the $\pm n$ Fourier channel pairs converge towards each other for smaller $\omega$, as expected from the fact that the energy differences between the channels will be smaller. 

\subsubsection{Phase dependence in scattering} \label{sec:phase_dep}
\begin{figure}
	\includegraphics[width=\linewidth]{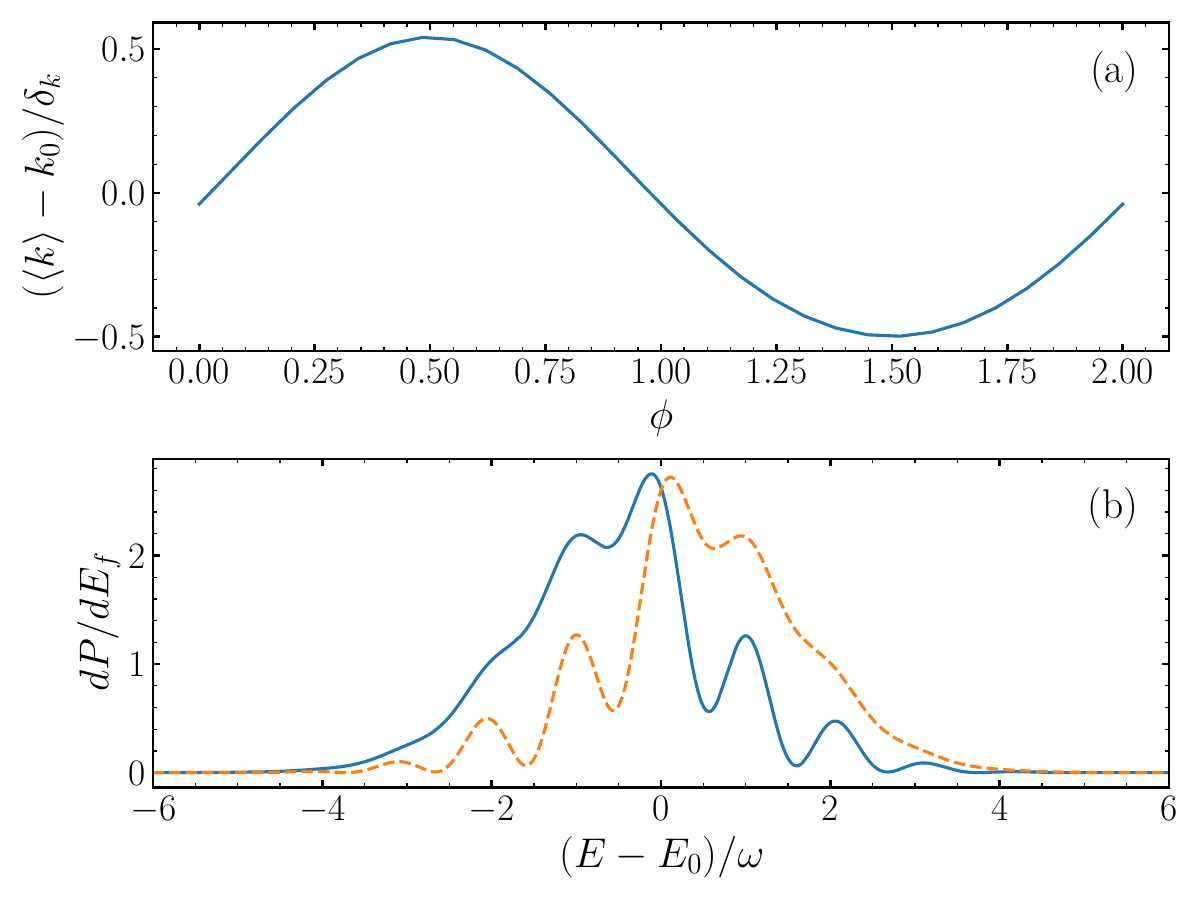}
	\caption{Results displaying a sensitivity to the phase, $\phi$ of the potential, when scattering with the wide $\sigma_k=0.007$ a.u. wave packet of Fig.~\ref{fig:k_dist_wide_wp}. (a) Mean momentum value of differential scattering probabilities for simulations with varying phase, $\phi$ of the potential. A clear sinusoidal behavior is observed. (b) Examples of two differential probabilities used to calculate panel (a), corresponding to those with maximum absolute deviation from $k_0$, $\phi \sim 1.5$ (full, blue) and $\phi\sim 0.5$ (dashed, orange).}
	\label{fig:phase_dependence}
\end{figure}

In this section, we briefly illustrate the curious result that our wave packet scattering theory can have a dependence on the phase, $\phi$, of the scattering potential Eq.~\eqref{explicit_hamiltonian}. Normally, when working with Floquet theory, such a dependence is not present as the interaction under consideration is assumed to span many (infinite) cycles \cite{joachain_atoms_2011}. In our case, however, the spatial localization of both the wave packet and the potential makes sub-cycle interactions possible, while our Hamiltonian remains periodic. 

We note that such results have been reported for wave packet scattering before in one dimensional calculations \cite{lefebvre_scattering_2005}. Here, it was argued that the required condition for phase sensitivity is that the phases of the different T-matrix elements can interfere. In our case, looking at Eq.~\eqref{diff_scatt_spherical} and Eq.~\eqref{gamma_factor}, we see that the possibility for the S-matrix elements to interfere is controlled by the wave packet width in momentum (energy), as given by $g(k)$. Assume, e.g., that the wave packet is tightly focused around a given $k_0$, such that the width of the wave packet in $k$-space, $\sigma_k \ll \omega/k_0$. Then $g(k_{in}) g^*(k_{in'}) \sim |g(k_{in})|^2\delta_{nn'}$, and consequently Eq.~\eqref{diff_scatt_spherical} reduces to 
\begin{multline}
	P_S(k_f) \sim \frac{1}{k_f} \sum_{\ell m} |c_{\ell m}|^2 
	\\
	\times \sum_n k_{in} |g(k_{in})|^2 \left|\delta_{n,0} - S^{(\ell m)}_{n0}(k_{in}) \right|^2
\end{multline}
and clearly, the Fourier channels no longer interfere. As already discussed, a wave packet like Eq.~\eqref{init_wp_spherical_seperated}, has an approximate temporal duration of $\tau \sim \sqrt{2\ln(2)}/\sigma_k k_0$. Together with the condition above, this means $\tau \gg 1/\omega$, and we see that interference between Fourier channels is lost when the duration of the wave packet is larger than the period of the potential oscillation, exactly as expected. 

Fig.~\ref{fig:phase_dependence} illustrates these points with the $\sigma_k=0.007$ a.u. wave packet from Fig.~\ref{fig:k_dist_wide_wp}~(a). This wave packet is wide enough in momentum that interference between different Floquet channels is possible. Correspondingly, its duration $\tau \sim 62$ a.u. ($1.5$ fs) is around half the period of the potential $T=110$ a.u. (2.7 fs) and we expect that some phase dependence should be present. In Fig.~\ref{fig:phase_dependence}~(a), the mean momentum value of the differential scattering probability is determined for simulations with various phases $\phi$. A clear sinusoidal modulation is present. Though not shown, a similar scan for the $\sigma_k=0.0005$ a.u. (21 fs) wave packet shows no dependence on $\phi$. Fig.~\ref{fig:phase_dependence}~(b) shows the two momentum distributions for values of $\phi$ that yielded the largest deviation in the average momentum, i.e. $\phi \sim 1.5$ (full, blue) and $\phi\sim 0.5$ (dashed, orange). Unsurprisingly, the two distributions are similar, one almost being the mirror image around the central energy, $E_0$, of the other. Each distribution has an increase in differential probability at energies to one side of $E_0$, and a decrease in differential probability to the other. Both panels of Fig.~\eqref{fig:phase_dependence} thus illustrates that, for temporally short wave packets, the exchange of Fourier energy quanta can be steered towards predominantly absorption or emission, depending on the phase, $\phi$. As the effect originates from interference terms,  it could be expected that a wave packet with even larger $\sigma_k$ (smaller duration, $\tau$) should have more possibility for interference, and thus potentially a larger sensitivity to $\phi$. An additional calculation for $\sigma_k=0.01$ a.u. (not shown) was found to support this expectation.

\subsection{Results with eikonal approximation} \label{sec:eikonal_res}
The parameters used in the eikonal simulations are identical to those from the previous section, unless otherwise stated. The eikonal functions $\phi_n$, needed in Eq.~\eqref{final_eikonal_scatt_prop} and Eq.~\eqref{final_eikonal_interference}, are found by solving the coupled differential equations in Eq.~\eqref{eikonal_diff_eqn} using a standard adaptive Runge-Kutta solver. Note that a new set of equations has to be solved for each value of $\bm r_\perp$. 

\begin{figure}
	\includegraphics[width=\linewidth]{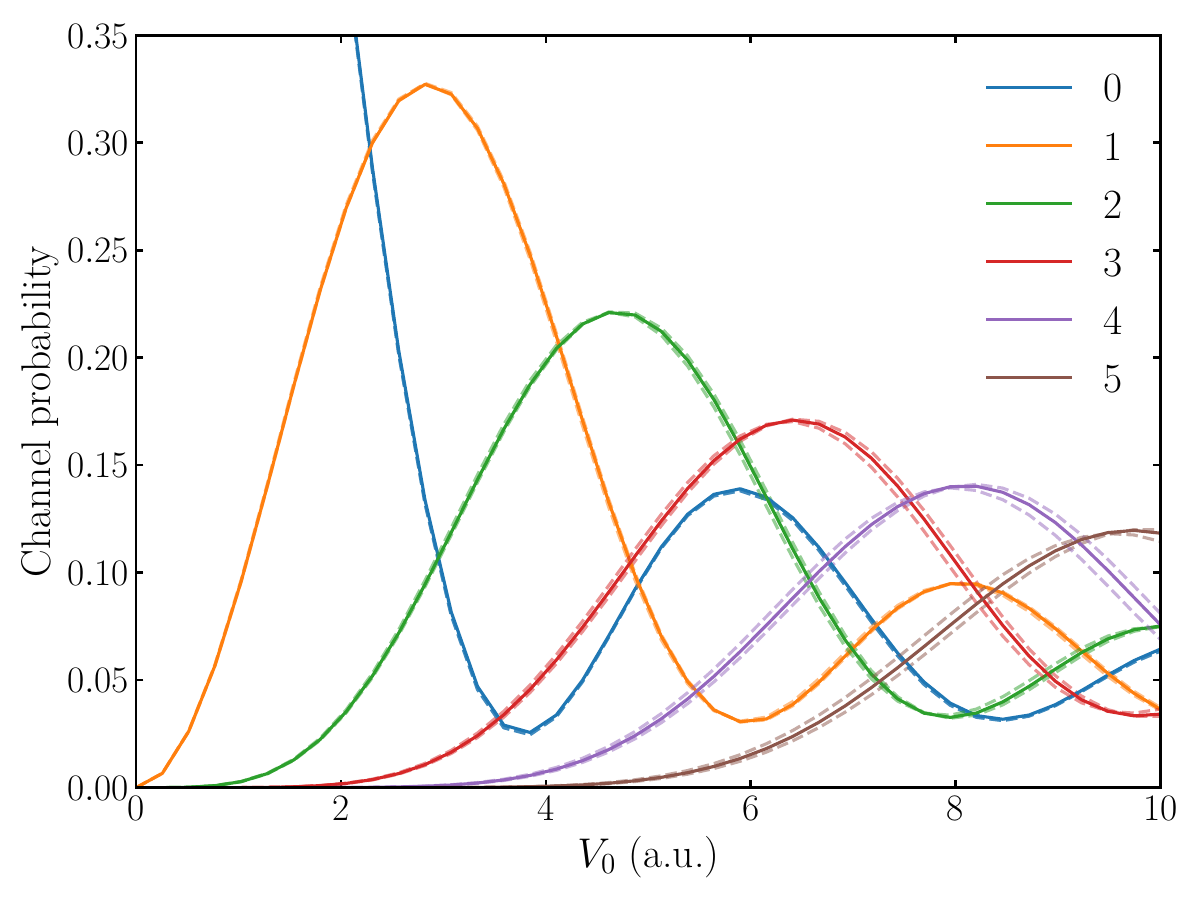}
	\caption{Comparison of Fourier channel probabilities obtained with the eikonal expressions Eq.~\eqref{final_eikonal_scatt_prop} and Eq.~\eqref{final_eikonal_interference} (solid lines) and those obtained with the R-matrix method from Fig.~\ref{fig:channel_prop_V_scan}~(d) (dashed lines). The color of the lines indicates the Fourier channel, $n$, as given by the legend. The eikonal results for the $\pm n$ channel pairs are identical. Note that the R-matrix results for $n=0,1$ are almost hidden behind the eikonal results on the scale of the figure.}
	\label{fig:Eikonal_V0_test}
\end{figure}
In the derivation of the eikonal results of Sec.~\ref{sec:eikonal_scatt_prop}, we assumed that the initial wave packet was separable in parallel and perpendicular components (both independently normalized). In practical calculations, we take the perpendicular wave packet to be a Gaussian 
\begin{equation} \label{wp_perp_gauss}
	\psi_{0\perp}(\bm k_\perp) = \frac{1}{\sqrt{2\pi \sigma_\perp^2}} \exp\left(-\frac{k_\perp^2}{4\sigma_\perp^2}\right),
\end{equation}
which has a real space transverse probability density, Eq.~\eqref{eikonal_perp_density},
\begin{equation} \label{gauss_trans}
\rho_\perp(\bm r_\perp) = 8\pi \sigma_\perp^2 e^{-2\sigma_\perp^2 r_\perp^2}.	
\end{equation}
We pick $\sigma_\perp = k_0 \sigma_\theta$ such that Eq.~\eqref{wp_perp_gauss} approximates the angular part of the $|k|$-Gauss wave packet, Eq.~\eqref{k-gauss}, for small $\theta$. 

For the scattering parameters used in Sec.~\ref{sec:R_mat_results}, the eikonal approximation for the Fourier channel probabilities, Eq.~\eqref{tot_eikonal_channel_prop}, is quite accurate. This is illustrated in Fig.~\ref{fig:Eikonal_V0_test}, where the eikonal results are compared to the channel probabilities of Fig.~\ref{fig:channel_prop_V_scan}~(d), obtained by the R-matrix approach. The eikonal results for the Fourier channel pairs $\pm n$ is identical, and thus only the positive $n$ results are shown. As seen, the eikonal results for the $n$'th Floquet channel tend to lie in between those obtained with the R-matrix method for the $\pm n$ channels. As already discussed, the R-matrix results for the $\pm n$ channels are close to each other for lower values of $|n|$, meaning the eikonal approximation works very well for the lowest channels. For higher channels and larger deviations in energy, the linearization in Eq.~\eqref{linear_k_diff_eikonal} becomes worse, and thus the R-matrix results start to differ. We note that a similar level of agreement is found between the eikonal approximation and R-matrix method for the $\sigma_\theta=0.02$ a.u. wave packet. 

In light of its accuracy in the present case, the eikonal approximation offers a different perspective on the strong vs. weak modulation discussed in terms of angular components in Fig.~\ref{sec:R_mat_res_mod_of_wp}. As discussed in Sec.~\ref{sec:sys_params}, a larger value of $\sigma_\theta$ corresponds to a smaller transversal width in real space $\sigma_{r_\perp}$. From Eqs.~\eqref{final_eikonal_scatt_prop}-\eqref{final_eikonal_interference}, this means that a smaller cross section of eikonal trajectories will contribute to the scattering. For the angular focused wave packet of  Fig.~\ref{fig:k_dist_narrow_wp}~(a), $\sigma_\theta=0.002$ a.u., the transversal width was comparable to that of the potential. In this case, the modulation is `lost' due to transversal averaging over many eikonal trajectories of the potential. Strong modulation, on the other hand, is achieved when this averaging is limited to a subset of the eikonal trajectories, corresponding to the case of Fig.~\ref{fig:k_dist_narrow_wp}~(b).

\section{Conclusion and outlook} \label{sec:conclusion_outlook}
In summary, we have derived a rigorous scattering theory for electronic wave packets interacting with oscillating, short-range potentials. By mapping the time dynamics into an extended Floquet space, the theory connects to time-independent multichannel plane wave scattering theory, allowing practical evaluation of scattering amplitudes. Calculation of such amplitudes was performed using an R-matrix approach, yielding semi-analytical results suited for practical calculations. Furthermore, it was demonstrated how a multichannel eikonal approximation leads to scattering probabilities similar to those of PINEM theory, with the addition of a weighting factor determined by the wave packet's transverse profile. 

The application of the theory was illustrated through a simple example: scattering on a spherically symmetric, time-oscillating potential. The results illustrated modulation of the wave packet into energy side bands, separated by the angular frequency of the potential. Additionally, the angular focus, or the transverse width, of the wave packet was seen to greatly affect the modulation, i.e., the distribution of probability in the different Fourier channels, with a focused wave packet yielding a stronger modulation. Finally, the eikonal approximation was tested against the R-matrix calculations, showing good agreement for the parameters considered.  

The presented theory integrates the modulation of electron wave packets into existing 3D wave packet scattering frameworks. The primary application is as a theoretical framework, allowing the calculation of modulated wave packets that can be used as input in other calculations, e.g. scattering on simple atomic targets. This will allow studies of how the modulation of wave packets can effect the scattering within a rigorous unified 3D framework. In order to transition to a realistic modulation with optical near-fields, the calculation of scattering amplitudes must be extended beyond the assumption of spherical symmetry and to larger spatial extents of the modulation potential. Another interesting modification of the presented theory would be generalization to scattering on quantized near-fields. In this case, the time-dependent semi-classical Hamiltonian is replaced by a time-independent one, effectively replacing the extended Floquet space with the Fock space of the quantized field.

\acknowledgments
We thank Dr. Yuya Morimoto for stimulating discussions.   This work is supported by the Independent Research Fund Denmark (Natural Sciences 10.46540/4283-00004B). 

\clearpage
\onecolumngrid
\appendix

\section{Time-dependent scattering}  \label{sec:app_time_dep_scatt}
In this appendix we show how to obtain Eq.~\eqref{time_dep_final}, following a regularization procedure from \cite{goldberger_collision}. First, we insert Eq.~\eqref{U_Floquet} in Eq.~\eqref{psi_time_evolv_expr}, along with a Fourier decomposition of the potential, to obtain
\begin{equation} \label{time_dep_scatt_eqn_expanded}
	|\psi(t)\rangle_\spaceIndex{A} = |\psi_\text{in}(t)\rangle_\spaceIndex{A} - i \lim\limits_{t_0 \to -\infty} \sum_m \sum_n \int_{t_0}^t dt' e^{in\omega t} {}_\spaceIndex{F}\langle n| e^{-i\hat{H}_\spaceIndex{FA}(t-t')} |0\rangle_\spaceIndex{F} e^{im\omega t'} \hat{V}_{\spaceIndex{A}m} e^{-i\hat{H}_{\spaceIndex{A}0}t'} |\psi_\text{in}\rangle_\spaceIndex{A}. 
\end{equation}
To handle the Hamiltonians, we write the in-state in a momentum space representation, Eq.~\eqref{psi_in_time_formalism}, and resolve the identity in eigenstates of the extended space Hamiltonian 
\begin{equation*}
	1_\spaceIndex{FA} = \sum_j |\phi_j\rangle_\spaceIndex{FA} \langle \phi_j |_\spaceIndex{FA}, \quad \hat{H}_\spaceIndex{FA} |\phi_j\rangle_\spaceIndex{FA} = E_j |\phi_j\rangle_\spaceIndex{FA}.
\end{equation*}
Doing this yields the following time integral 
\begin{equation}
	I_t = \int_{t_0}^t dt' \, \exp[it'(E_j + m\omega - E_k)] = \frac{-i}{E_j +m\omega - E_k} \left( \exp[it(E_j + m\omega - E_k)] - \exp[it_0(E_j + m\omega - E_k)] \right).
\end{equation}
While the term corresponding to the upper limit provides no difficulties, the term from the lower limit is not well-defined in the $t_0 \to -\infty$ limit. We handle this by introducing a complex term $i\eta$ as $I_t = \lim\limits_{\eta \to 0} I_t^+(\eta)$, with 
\begin{align} \label{I_reg_expanded}
	I_t^+(\eta) &=  \frac{-i}{E_j - i\eta + m\omega - E_k} \left( \exp[it(E_j + m\omega - E_k) +\eta t] - \exp[it_0(E_j + m\omega - E_k) +\eta t_0] \right) \nonumber 
	\\ 
	&= \frac{-i\exp[it(E_j + m\omega - E_k) +\eta t]}{E_j - i\eta + m\omega - E_k} - \int_{-\infty}^{t_0} d\tau \exp[i\tau(E_j + m\omega - E_k) +\eta \tau],
\end{align}
where the last rewriting is allowed by the regularization from the $\eta\tau$ term. Consider the role of the last term above, in the original Eq.~\eqref{time_dep_scatt_eqn_expanded}. This corresponds to terms with $\hat{V}(\tau)|\psi_\text{in}(\tau)\rangle_\spaceIndex{A}$, which by Eq.~\eqref{wp_finite_condition} can be taken as zero for $t_0<t_c$. As we are interested in the case $t_0 \to -\infty$, we thus see that the second term in Eq.~\eqref{I_reg_expanded} has no contribution in our final scattering state. In the end Eq.~\eqref{time_dep_scatt_eqn_expanded} becomes 
\begin{align}
	|\psi(t)\rangle_\spaceIndex{A} &= |\psi_\text{in}(t)\rangle_\spaceIndex{A} - \lim\limits_{\eta \to 0} \sum_m \sum_n \int d \bm k \, {}_\spaceIndex{F}\langle n| e^{in\omega t} \sum_j \frac{\exp[-it(E_k - m\omega+i\eta)]}{E_j -i\eta + m\omega - E_k}  |\phi_j\rangle \langle \phi_j | \hat{V}_{\spaceIndex{A}m} \psi_0(\bm k) |0\rangle_\spaceIndex{F} | \bm k \rangle_\spaceIndex{A} \nonumber
	\\ 
	&= \lim\limits_{\eta \to 0}\int d\bm k \, \psi_0(\bm k) e^{-iE_kt} \left( |\bm k \rangle_\spaceIndex{A} -  \sum_n e^{in\omega t} \sum_m e^{im\omega t + \eta t} {}_\spaceIndex{F}\langle n| (\hat{H}_\spaceIndex{FA} - i\eta + m\omega - E_k)^{-1} |0\rangle_\spaceIndex{F} \hat{V}_{\spaceIndex{A}m} | \bm k \rangle_\spaceIndex{A} \right) \nonumber
	\\ 
	&= \lim\limits_{\eta \to 0}\int d\bm k \, \psi_0(\bm k) e^{-iE_kt} \left( |\bm k \rangle_\spaceIndex{A} + \sum_n e^{in\omega t} \sum_m e^{im\omega t + \eta t} \hat{\mathcal{G}}^{(+)}_{n0}(E_k - m\omega) \hat{V}_{\spaceIndex{A}m} | \bm k \rangle_\spaceIndex{A} \right), 
\end{align}
thus obtaining Eq.~\eqref{time_dep_final} of the main text.  

\section{Asymptotic states in momentum space} \label{sec:app_asymp_state_momentum}
Often we are interested in the asymptotic scattering wave function in momentum space. Instead of the real-space condition $r \to \infty$, which lead to spherical waves in Eq.~\eqref{real_space_scatt_wf_asymp}, we in momentum space consider $t \to \infty$ leading to energy conservation. To see how this comes about, consider the scattering wave function, Eq.~\eqref{floquet_scatt_wavefunc}, in momentum space 
\begin{align} \label{initial_momentum_space_expr}
	\langle \bm k_f |\psi(t)\rangle_\spaceIndex{A} &= \psi_0(\bm k_f) e^{-iE_{k_f}t} + \sum_n \int d \bm k \, \psi_0(\bm k) e^{-iE_k t + in\omega t} (E_k - E_{k_f} - n\omega + i\eta)^{-1} \sum_m \langle \bm k_f | \hat{V}_{nm} | \psi^{(+)}_{m,\bm k} \rangle_\spaceIndex{A} \nonumber
	\\ 
	&= \psi_0(\bm k_f) e^{-iE_{k_f}t} -\frac{1}{4\pi^2} \sum_n \int d\bm k \, \psi_0(\bm k) e^{-iE_k t + in\omega t} (E_k - E_{k_f} - n\omega + i\eta)^{-1} f_n(\bm k_f, \bm k). 
\end{align}
Here the first equality follows using Eq.~\eqref{LS_floquet} when taking $| \psi_0\rangle_\spaceIndex{FA} = |0\rangle_F |\bm k\rangle_A$, and the second equality follows from the relation between $T$-matrix elements and the scattering amplitude, Eq.~\eqref{scatt_amp_vs_Tmat}. Note that the scattering amplitude in this expression is evaluated in arbitrary $\bm k_f$, $\bm k$, i.e., no energy conservation has been imposed yet.  

Asymptotically, as $t\to\infty$, we can evaluate the integral over $k$ with the methods of Appendix~\ref{sec:asymp_int_trick}. Switching variables to energy, $E_k=k^2/2$, the relevant integral of Eq.~\eqref{initial_momentum_space_expr} becomes 
\begin{align} \label{app_to_k_space_int}
	I_E &= \int_0^\infty dE_k \sqrt{2E_k}  \psi_0(\sqrt{2E_k}\hat{\bm k})e^{-iE_kt}(E_k-E_{k_f}-n\omega + i\eta)^{-1} f_n(\bm k_f, \sqrt{2E_k}\hat{\bm k}) \nonumber
	\\
	&\xrightarrow[t \to \infty]{} -2\pi i k_{in} \psi_0(k_{in}\hat{\bm k})f_n(\bm k_f, k_{in}\hat{\bm k})e^{-iE_{k_{in}}},
\end{align}
where $E_{k_{in}} = E_{k_f} + n\omega$ is the initial energy, such that the final energy is $E_{k_f}$ under emission of $n$ Fourier energy quanta $\omega$. $k_{in}=\sqrt{2E_{k_{in}}}$ is the corresponding momenta. Using Eq.~\eqref{app_to_k_space_int} in Eq.~\eqref{initial_momentum_space_expr} yields the final expression for the asymptotic scattering wave function in momentum space  
\begin{equation}
	\psi(\bm k_f) \xrightarrow[t \to \infty]{} e^{-iE_{k_f}t} \left[ \psi_0(\bm k_f)  + \frac{i}{2\pi} \sum_n \int d \Omega_k k_{in} \psi_0(k_{in}\hat{\bm k}) f_n(\bm k_f, k_{in}\hat{\bm k}) \right],
\end{equation}
as stated in Eq.~\eqref{scatt_state_momentum} of the main text.

\section{Asymptotic limit of Fourier-like integral}  \label{sec:asymp_int_trick}
Let us consider integrals of the form 
\begin{equation}
	I = \int_0^\infty dx\, \frac{f(x)e^{-ixt}}{x-x_0 + i\eta}.
\end{equation} 
We wish to evaluate this integral in the $t \to \infty$ limit. In the following we assume that $f(x)$ is finite everywhere. First, we split the integral into 3 intervals: 
\begin{equation} \label{app_I_split}
	I = \int_0^{x_0-\epsilon} dx\, \dots + \int_{x_0-\epsilon}^{x_0+\epsilon}dx\, \dots + \int_{x_0+\epsilon}^{\infty}dx\, \dots.
\end{equation}
Clearly, the two outer integrals (the integrals over $[0,x_0-\epsilon]$ and $[x_0+\epsilon,\infty[$), are well behaved, even in the $\eta \to 0$ limit, as the effect of the pole is isolated to the central integral. The contribution of the outer integrals as $t \to \infty$ can be found using standard integration by parts \cite{bleistein_asymptotic_1986}. Consider 
\begin{equation}
	\int_0^a dx \, g(x) e^{-ixt} = \int_0^a dx\, \frac{g(x)}{-it} \frac{d}{dx} e^{-ixt} = \frac{g(x)}{-it}e^{-ixt}\bigg\rvert_0^a - \int_0^a dx\, \frac{g'(x)}{-it} e^{-xit}.
\end{equation}
Assuming higher-order derivatives of $g(x)$ are well-defined, we can use the same integration by parts procedure on the remaining integral, such that 
\begin{equation}
	\int_0^a dx \, g(x) e^{-ixt} = \frac{g(x)}{-it}e^{-ixt}\bigg\rvert_0^a + \mathcal{O}(1/t^2)
\end{equation}
For $t\to \infty$ we thus have, as long as $g$ remains finite on the interval of interest, 
\begin{equation} \label{app_limit_outer_int}
	\int_0^a dx \, g(x) e^{-ixt} \to 0, \quad t\to \infty.
\end{equation}
Returning to Eq.~\eqref{app_I_split}, we can apply Eq.~\eqref{app_limit_outer_int} to the outer integrals. The central integral (over $[x_0-\epsilon, x_0+\epsilon´]$) requires a little more attention in the $\eta \to 0$ limit.
\begin{equation}
	I \xrightarrow[t \to \infty]{} \int_{x_0-\epsilon}^{x_0+\epsilon} dx \, \frac{f(x)e^{-ixt}}{x-x_0 + i\eta}.
\end{equation}
We handle this integral by contour integration, picking a semi-circular contour in the lower complex plane, enclosing the pole. Normally, the radius of this semi-circular contour is taken to be $R\to\infty$ in order to employ Jordan's lemma \cite{arfken_mathematical_2013}, but in this case the contour can remain at a small finite radius, as we consider $t\to\infty$ instead. We note that this works even when $f(x)$ diverges in the imaginary directions of the complex plane. From the residue theorem \cite{arfken_mathematical_2013} we finally obtain,
\begin{equation}
	I \xrightarrow[t \to \infty]{} -2\pi i f(x_0-i\eta) e^{-i(x_0-i\eta)t}.
\end{equation} 
Note that if we do not consider $\eta\to 0$ the resulting integral is 0, as is expected based on the integration by parts procedure above.

\section{The R-matrix method} \label{sec:app_r_mat_method}
This appendix covers the details of the R-matrix method used to calculate the scattering amplitudes behind the numerical results of Sec.~\ref{sec:R_mat_results}. While much material about this method can be found in the literature, see Ref.~\cite{burke_rmat_2011} and references herein, we include this appendix to make it approacable for the reader, in the notation of this paper. 

In the R-matrix approach, space is separated into an inner and outer part. In the inner part the scattering is treated by exact diagonalization (within a given basis choice). The resulting eignvectors are used to create a solution with the correct asymptotic boundary conditions by a matching procedure. Usually the R-matrix method is used to separate the region of space where exchange effects are important from the rest. In our case, we employ the method to ensure correct matching of the different Fourier channels, and also to easily allow calculation of scattering amplitudes for several energies, as required by Eq.~\eqref{scatt_state_momentum}. 

The R-matrix method is readily formulated in spherical coordinates $(r,\theta,\phi)$. To obtain radial equations, we expand the scattering states as 
\begin{equation}
	\langle \bm r|\psi^{(+)}\rangle_\spaceIndex{FA} = \sum_n \sum_{\ell, m} |n\rangle_\spaceIndex{F} \frac{u_{n \ell m}(r)}{r} Y_\ell^m(\hat{\bm r}). 
\end{equation}
Inserting in Eq.~\eqref{coupled_floquet_eqns} and projecting down onto $Y_\ell^{m*}$, we obtain
\begin{equation} \label{coupled_rad_equations}
	\frac{1}{2}\frac{du_{n\ell m}}{dr^2} + \left( E - n\omega - \frac{\ell(\ell+1)}{2r^2} \right) u_{n\ell m}(r) = \sum_p \int d\Omega \, Y_\ell^{*m} \hat{V}_{\spaceIndex{A}p} \sum_{\ell' m'} Y_{\ell'}^{m'}	u_{n-p,\ell' m'}(r).
\end{equation}
We take the inner region to be contained within the R-matrix boundary at $r=a_0$, and wish to solve Eq.~\eqref{coupled_rad_equations} within this sphere. Imposing this boundary, however, makes the radial kinetic energy operator non-hermitian. To fix this, we introduce the Bloch operator \cite{burke_rmat_2011}
\begin{equation}
	\hat{\mathcal{L}}_\spaceIndex{FA} = \frac{1}{2} \delta(r-a_0) \left(\frac{d}{dr}-\frac{b_0-1}{r} \right).
\end{equation}
Here $b_0$ is an arbitrary constant, which is usually set to 0. Adding to Eq.~\eqref{rmat_floquet_eqn}
\begin{equation} \label{rmat_floquet_bloch}
	(\hat{H}_\spaceIndex{FA} +  \hat{\mathcal{L}}_\spaceIndex{FA})|\psi^{(+)}\rangle_\spaceIndex{FA} = (E+\hat{\mathcal{L}}_\spaceIndex{FA})|\psi^{(+)}\rangle_\spaceIndex{FA},
\end{equation}
we instead have to diagonalize the operator $\hat{H}_\spaceIndex{FA} +  \hat{\mathcal{L}}_\spaceIndex{FA}$. Denoting the eigenstates of this operator $|\psi_k\rangle_\spaceIndex{FA}$, where $(\hat{H}_\spaceIndex{FA} +  \hat{\mathcal{L}}_\spaceIndex{FA})|\psi_k\rangle_\spaceIndex{FA} = E_k |\psi_k\rangle_\spaceIndex{FA}$, we can rewrite Eq.~\eqref{rmat_floquet_bloch} as 
\begin{equation} \label{rmat_floqet_spectral}
	|\psi^{(+)}\rangle_\spaceIndex{FA} = \frac{1}{\hat{H}_\spaceIndex{FA} +  \hat{\mathcal{L}}_\spaceIndex{FA} - E}  \hat{\mathcal{L}}_\spaceIndex{FA}|\psi^{(+)}\rangle_\spaceIndex{FA} = \sum_k |\psi_k\rangle_\spaceIndex{FA} \frac{1}{E_k - E}{}_\spaceIndex{FA}\langle\psi_k|  \hat{\mathcal{L}}_\spaceIndex{FA}|\psi^{(+)}\rangle_\spaceIndex{FA}.
\end{equation}
To obtain radial equations, we assume that both $|\psi_k\rangle_\spaceIndex{FA}$ and $|\psi^{(+)}\rangle_\spaceIndex{FA}$ can be expanded as 
\begin{equation}
	\langle \bm r|\psi^{(+)}\rangle_\spaceIndex{FA} = \sum_n \sum_{\ell, m} |n\rangle_\spaceIndex{F} \frac{u_{n \ell m}(r)}{r} Y_\ell^m(\hat{\bm r}). 
\end{equation}
Inserting in Eq.~\eqref{rmat_floqet_spectral} and projecting onto $|n\rangle_\spaceIndex{F}$ and $Y_\ell^{m*}$ we obtain 
\begin{equation}
	u_{nlm}(r) = \sum_{n', \ell', m'} \frac{1}{2a_0}\sum_k \frac{g^{(k)}_{n\ell m}(r)g^{(k)}_{n'\ell' m'}(a_0)}{E_k-E} \left(a_0\frac{du_{n'\ell' m'}}{dr'} - b_0 u_{m' \ell' m'}(r') \right)\bigg\rvert_{r'=a_0}.
\end{equation}
Here $g_{nlm}^{(k)}$ is the radial functions corresponding to the eigenstates $|\psi_k\rangle_{FA}$. Defining the R-matrix elements as 
\begin{equation} \label{rmat}
	R_{n \ell m, n'\ell' m'}(E) = \frac{1}{2a_0}\sum_k \frac{g^{(k)}_{n\ell m}(a_0)g^{(k)}_{n'\ell' m'}(a_0)}{E_k-E},
\end{equation}
the amplitude of the radial function for the scattering state on the R-matrix shell ($r=a_0$) can be written 
\begin{equation} \label{rmat_eqn_radial_eqn}
	u_{nlm}(a_0) = \sum_{n', \ell', m'} R_{n \ell m, n'\ell' m'}(E) \left(a_0\frac{du_{n'\ell' m'}}{dr'} - b_0 u_{m' \ell' m'}(r') \right)\bigg\rvert_{r'=a_0}.
\end{equation}
The procedure for calculation is now the following: First, the R-matrix elements are constructed by numerically diagonalizing the inner region per. Eq.~\eqref{rmat_floquet_bloch}, and assembling the matrix-elements using Eq.~\eqref{rmat}. The radial functions in Eq.~\eqref{rmat_eqn_radial_eqn} are then expressed as linear combination of suitable asymptotic forms, defining a system of linear equations that can be solved for the coefficients. The asymptotic forms are related to the scattering amplitude, which is hereby obtained. The following subsection outlines the details of this procedure.

We end by noting, that by far the most expensive step in determining the scattering amplitude through the above-mentioned procedure, is the diagonalization of the Floquet-Hamiltonian in Eq.~\eqref{rmat_floquet_bloch}. Once this has been done, the determination of scattering amplitudes at different energies only requires constructing the R-matrix elements Eq.~\eqref{rmat} and solving the linear system of Eq.~\eqref{rmat_eqn_radial_eqn}. 

\subsubsection{Matching to asymptotic conditions}
Outside the R-matrix boundary, we assume that the coupling potentials $\hat{V}_{\spaceIndex{A}p}$ are zero. We thereby obtain a set of standard uncoupled radial Schrödinger equations, the solution of which are known to be Riccati-Bessel functions \cite{joachain_collision}. As a basis to describe our asymptotic states we thus consider the following functions, describing scattering from channel $j\to i$ \cite{friedrich_theoretical_2017}
\begin{equation} \label{F_func_expr}
	F_{ij} \xrightarrow[r \to \infty]{} \sqrt{\frac{2}{\pi k_i}} \left( k_ir j_{\ell_i}(k_ir)\delta_{i,j} - K_{ij} k_i r n_{\ell_i}(k_ir) \right),
\end{equation} 
where $j_\ell$ and $n_\ell$ are spherical Bessel and Neumann functions respectively, $K_{ij}$ are expansion coefficients known as K-matrix elements and the front factor ensures normalization in energy. Note that $i,j$ are really compound indices, describing all quantum numbers that can change during the scattering. Defining 
\begin{equation}
	\phi_i^j = \sqrt{\frac{2}{\pi k_i}}k_i r j_{\ell_i}(k_i r), \quad \phi_i^n = -\sqrt{\frac{2}{\pi k_i}}k_i r n_{\ell_i}(k_i r),
\end{equation}
we write the total system system of $F_{ij}$ functions for scattering from and to all states, in the matrix form 
\begin{equation} \label{F_mat_expr}
	\bm F = \bm \phi^j + \bm \phi^n \bm K,
\end{equation}
using the matrix elements $[\bm \phi^{j/n}]_{ij} = \phi_i^{j/n} \delta_{ij}$. In order to match Eq.~\eqref{F_mat_expr} to the solutions in the inner R-matrix region, we evaluate $\bm F$ in $r=a_0$, and insert into Eq.~\eqref{rmat_eqn_radial_eqn}
\begin{equation}
	\bm F(a_0) = a_0 \bm R \dot{\bm F}(a_0), 
\end{equation}
where we have taken the arbitrary constant $b_0=0$. Inserting Eq.~\eqref{F_mat_expr} and rearranging, we obtain a matrix equation for the K-matrix 
\begin{equation}
	\left(\bm \phi^n(a_0) - a_0 \bm R \dot{\bm \phi^n}(a_0)\right) \bm K = -\bm \phi^j(a_0) + a_0 \bm R \dot{\bm \phi^j}(a_0),
\end{equation} 
which is easily solvable numerically. 

The form of our asymptotic functions in Eq.~\eqref{F_func_expr} is not directly relatable to the scattering amplitude, which is related to outgoing spherical waves, see Eq.~\eqref{real_space_scatt_wf_asymp}. A more suitable choice of basis functions might therefore be 
\begin{equation}
	G_{ij} \xrightarrow[r \to \infty]{} \sqrt{\frac{2}{\pi k_i}} \left(e^{-i(k_ir -\pi \ell_i /2)}\delta_{ij} - e^{i(k_ir -\pi \ell_i /2)}S_{ij}\right), 
\end{equation}
thus defining the S-matrix elements. The two choices are related in the $r \to \infty$ region, yielding the relation \cite{friedrich_theoretical_2017}
\begin{equation}
	\bm S = (\bm 1 + i \bm K)(\bm 1 - i \bm K)^{-1},
\end{equation}
meaning the $\bm S$ matrix is easily obtainable once the $\bm K$ matrix has been found. 

We are finally in position to link to the scattering amplitude. First, we write Eq.~\eqref{real_space_scatt_wf_asymp} in the more general form
\begin{equation} \label{real_space_scatt_wf_asymp_general}
	\psi^{(+)}_{n,n_i}(\bm r) \xrightarrow[r \to \infty]{} \frac{1}{(2\pi)^{3/2}} \left( e^{i \bm k_i \cdot \bm r}\delta_{n,n_i} + f_{n \leftarrow n_i}(k_n\hat{\bm r}, \bm k_i)\frac{e^{i\bm k_n r}}{r} \right),
\end{equation}
now describing scattering from a general initial state $n_i$ to final $n$. Expanding the scattering state in a sum of spherical harmonics, we write its radial function in terms of the $G_{ij}$ basis as 
\begin{equation}
	u_{n\ell m}^{(n_i)}(r) = \sum_{\ell_i m_i} A_{\ell_i m_i} G_{n\ell m, n_i \ell_i m_i}(r).
\end{equation}
To determine the expansion coefficients, $A_{\ell_i m_i}$, we expand the plane wave part of Eq.~\eqref{real_space_scatt_wf_asymp_general}, and compare the incoming wave parts. Forcing them to be identical means 
\begin{equation}
	A_{\ell m} = - \frac{1}{2\sqrt{k_n}} i^{\ell-1} Y_\ell^{m*}(\hat{\bm k_i}).
\end{equation}
Using this in the matching of the outgoing waves finally yields the relation between $f_{n\leftarrow n_i}$ and the S-matrix elements
\begin{equation} \label{app_scatt_amp_S_mat_relation_general}
	f_{n\leftarrow n_i} = \frac{2\pi i}{\sqrt{k_i k_n}} \sum_{\ell m} Y_\ell^m(\hat{\bm r}) \sum_{\ell_i m_i} i^{\ell_i - \ell} Y_{\ell_i}^{m_i}(\hat{\bm k_i})^* \left(\delta_{n,n_i}\delta_{\ell, \ell_i}\delta_{m,m_i} - S_{n\ell m, n_i \ell_i m_i}\right). 
\end{equation}
In the main text, the incoming state is always considered to be in the $n_i=0$ channel, so while the S-matrices for the different incoming channels are determined, they are not used in the results of the paper. Thus picking $n_i=0$ in Eq.~\eqref{app_scatt_amp_S_mat_relation_general} yields Eq.~\eqref{scatt_amp_S_mat_relation} of the main text, where we have used that asymptotically $\hat{\bm k}_f = \hat{\bm r}$, similar to Eq.~ \eqref{real_space_scatt_wf_asymp}. 

\twocolumngrid
\bibliography{sovs}

\end{document}